Chapter

# Perspective Chapter: MOOCs in India: Evolution, Innovation, Impact, and Roadmap

*Partha Pratim Das*

## Abstract

With the largest population of the world and one of the highest enrolments in higher education, India needs efficient and effective means to educate its learners. India started focusing on open and digital education in 1980's and its efforts were escalated in 2009 through the NMEICT program of the Government of India. A study by the Government and FICCI in 2014 noted that India cannot meet its educational needs just by capacity building in brick and mortar institutions. It was decided that ongoing MOOCs projects under the umbrella of NMEICT will be further strengthened over its second (2017–21) and third (2021–26) phases. NMEICT now steers NPTEL / SWAYAM (India's MOOCs) and several digital learning projects including Virtual Labs, e-Yantra, Spoken Tutorial, FOSSEE, and National Digital Library on India—the largest digital education library in the world. Further, India embraced its new National Education Policy in 2020 to strongly foster online education. In this chapter, we take a deep look into the evolution of MOOCs in India, its innovations, its current status and impact, and the roadmap for the next decade to address its challenges and grow. AI-powered MOOCs is an emerging opportunity for India to lead MOOCs worldwide.

## 1. Introduction

> *MOOCs in India have evolved significantly over the past decade, providing access to quality education, bridging the skills gap, and democratizing learning, with a future focused on localized content, emerging technologies, and partnerships with employers.*
>
> – ChatGPT, 23:22, 14 May 2023.

India is a young nation in an aging world. The median age in India is about 28, compared to 37 in China and the US, 45 in Western Europe, and

Massive Open Online Courses - Current Practice and Future Trends

49 in Japan. This simple stats, known as *demographic dividend*[1], is the main steam behind the resurgence and emergence of India as a global leader in various domains. Early this year (2023), India has surpassed China as the world's most populous country [1] and as one of the largest work-forces[2] globally. During the next 25 years, one in five working-age group persons will be living in India. As ET rightly quotes [2] a report of *Confederation of Indian Industry* (CII)

> *"If India's demographic dividend is productively employed, growth prospects will brighten, helping it to leapfrog its GDP from the current $3 trillion to $9 trillion by 2030 and $40 trillion by 2047."*

This certainly is an extraordinary opportunity with a temporal window spanning nearly five decades [3] from 2005 to 2006 to 2055–2056. India will need to focus on the right policies [4] to reap rich dividends from this demographic transition.

The mammoth size of the work force alone cannot take the nation forward; in fact it may become a huge liability, unless this working age population is skilled and is employable. According to *India Skill Reports 2022* and *2023* [5–7], employability in India has improved[3] from 33% in 2014 to 46.2% in 2022, and to 50.2% in 2023. Yet, nearly half of the working population is still unemployable. This certainly is a big concern.

Raising education standards, in both rural and urban sectors, is naturally of foremost priority [3]. Every child needs to complete high school and go on to skilling, training, and vocational education. Only schools with modern curricula, open digital universities for higher education, and adoption of *Massive Open Online Courses* (MOOCs) at scale can contribute significantly to India's qualified workforce.

Thoughts about scaling the quantity as well as quality of education has been around since the onset of the demographic transition window[4]. However, in 2013, *Ministry of Education*[5] (MoE) undertook an in-depth study on the issue jointly with *Federation of Indian Chambers of Commerce & Industry* (FICCI). The findings firmly corroborated the adoption of MOOCs at a national level. We quote:

---

[1] Demographic Dividend is *the economic growth potential that can result from shifts in a population's age structure, mainly when the share of the* working-age population (15–64) *is larger than the non-working-age share of the population (< 14 and > 65)*. https://www.unfpa.org/demographic-dividend

[2] 25% population is in the age group of 0–14 years, 18% in 10–19 years, 26% in 10–24 years, and *68% in* 15–64 *years*, and only 7% above 65 years [1].

[3] This quoted figure is for employability of men. It is 52.80% for women in 2023. Interestingly, India has had higher employability for women for many years

[4] Ministry of Education (MoE) launched several programs – NPTEL in 2001, e-Yantra in 2003, and NMEICT & Virtual Lab in 2009. We discuss in depth in Section 2

[5] Erstwhile Ministry of Human Resource Development (MHRD) from 1985 to 2020

> *"The Indian higher education system has undergone massive expansion to become the largest in the world enrolling over 70 million students. Such expansion would have been unimaginable without the extensive use of Information and Communication Technology (ICT) tools. To illustrate, if India were to create this additional capacity through increase in brick and mortar institutions alone, it would have had to build six universities and 270 colleges each and every month in the last 20 years have been impossible to achieve with India's limited resources. Instead, India chose to go the MOOCs way."* **Source:** *From 2013 to 2030: the nuts and bolts of transformation* [8], p. 11. More in [9]

This reaffirmation of faith on MOOCs resulted in accelerated growth of the digital education system in India riding on the wings of *National program on Technology Enhanced Learning* (NPTEL) from 2001, *NPTEL Online Certification* (NPTEL-NOC) from 2014, *Study Webs of Active-Learning for Young Aspiring Minds* (SWAYAM, India's own MOOCs) from 2016, *University Grants Commission's* (UGC) approval of credit earning from SWAYAM in 2016, and finally *National Education Policy* (NEP) in 2020.

The pandemic put a brake on the world and the world leaders of MOOCs like *Coursera*, *edX*, and *FutureLearn* lost 30 to 50% of enrolment [10] in 2021 over 2020. But SWAYAM managed to stay at the same level as India moved faster on MOOCs to protect and build its education. And since then it has been growing strongly—not only in numbers, but also in depth, diversity, and deployment. We tell that story in this chapter.

## 1.1 Background

Starting with *Vedic Education* ($\sim$1500 BCE) and subsequent *Buddhist Education* ($\sim$500 BCE) India has had a rich tradition of education systems since the Ancient era. It followed the *Gurukul system of learning* where the student (*shishya*) lived near or with the teacher (*guru*) and got educated through oral instructions (*shruti*) in an *one-on-one interaction*. Over time the need of scale led to the emergence of education hubs[6] [11] at multiple places such as Takshashila ($\sim$600 BCE), Nalanda ($\sim$400 CE), and Vikramashila ($\sim$800 CE). These institutes continued with the gurukul style and drew learners from far and middle east. However, *scriptures* (written forms of education) started to complement *shruti* and *group-instructions* started to supplement one-on-one instructions. Education systems continued flourish in the Medieval ($\sim$600 CE–$\sim$1600 CE) and Early Modern (Moghul) eras (1526 CE–1850 CE) through the proliferation [11] of *Maktabs* and *Madrasas*[7] at places such as Delhi, Agra, Jaunpur, and Bidar. Beginning of the Modern era witnessed the invasion of and capture by the British. They systematically dismantled[8] [12] the traditional Indian education system[9] to introduce classrooms-centric teaching with multitudes of rigid institutional

---

[6] Much like the universities today

[7] Owing to the dominance of the Muslim rulers

[8] *"I propose that we replace her (India's) old and ancient education system, her culture, for if the Indians think that all that is foreign and English is good and greater than their own, they will lose their self-esteem, their native culture and they will become what we want them, a truly dominated nation"* – Lord Macaulay's Address to the British Parliament on February 2, 1835

[9] *"The beautiful tree of education was cut down by you British. Therefore today India is far more illiterate than it was 100 years ago."*—Mahatma Gandhi, Round-table conference in 1931

framework. For over 150 years India has been following that system to the letters. In this chapter, we refer to this as *Traditional* or *Brick-and-Mortar*[10] or *Classroom System of Education*.

Since independence in 1947, India focused on building educational institutions[11] at national scale. While many of these institutions such as IITs, IIMs, AIIMS started to shine very soon, within a few decades the enormity of the scale of building brick-and-mortar institutions appeared daunting and India launched its *Open and Distance Learning*[12] (ODL) programs through *Indira Gandhi National Open University* [13, 14] (IGNOU).

Easy availability of *personal computers* brought in a paradigm shift in the way students can learn using technology and *Digital Learning*[13] started becoming commonplace in the 1990s. Teachers and students started using technology in the forms of soft copies of books, presentations, audios, videos, contents from repositories. During the early years, digital learning used technology yet the teacher and the student continued to be co-located at (almost) all times.

Advent of the *internet* elevated digital learning to the next level where the teacher and the student could be separated in space in *Remote/Online Learning*[14]. The education is relayed through technology, such as discussion boards, video conferencing, and online assessments.

Finally, online learning took to a massive scale with the introduction of *MOOCs* worldwide. We discuss that in the next section in depth as understanding of the MOOCs is key to understanding this chapter. In **Table 1**, we summarize the discussions above with a brief overview for and ontology of education systems in India.

### 1.1.1 Massive open online education (MOOCs)

*MOOC*s is ***Massive*** (designed to work at wide scales—from 10 to 10 thousand students or more), ***Open*** (for registration to all, for access to any content, and for free or affordable education), ***Online*** (using access to local or remote cohorts and real-time interactions over internet), ***Courses*** (self-paced with scripted assignments and instructor feedback—preferably with credit mobility and learning community support). MOOCs originated from *connectivist cMOOCs* (with focus on community and connections), and *extended xMOOCs* (with focus on scalability) in the early years of the millennium with the term "MOOC"[15] being coined in 2008. Many US universities started offering free online courses for the public and *Coursera* [16] (Stanford), *edX* [17]

---

[10] Brick-and-Mortar education occurs at a physical school, as opposed to a virtual environment

[11] IITs (1950), RECs (1959), ITIs (1950), IIMs (1961), AIIMS (1956), Law & Medical Colleges, Kendriya Vidyalaya (1962), etc.

[12] ODL is a system wherein teachers and learners need not necessarily be present either at same place or same time and is flexible in regard to modalities and timing of teaching and learning as also the admission criteria without compromising necessary quality considerations.

[13] Digital learning is defined as any type of learning accompanied by the use of technology, or instructional practice that makes use of technology

[14] In Remote Learning the student and the educator, or information source, are not physically present in a traditional classroom environment

[15] Mishra [15] points to the Clark taxonomy of variants of MOOCs.

| Name of learning paradigms | Mode of study material | Space: teacher and student | Time: Sync/ async | Intervening medium | Mode of interaction |
|---|---|---|---|---|---|
| *Gurukul* | Physical | Co-located | Sync | Face-to-face | One-to-One |
| *Classroom* | Physical | Co-located | Sync | Face-to-face | One-to-Many |
| *Open & Distance* | Physical | Separated | Async | Postal Mail | Sparse |
| *Digital* | Electronic | Co-located | Sync | Face-to-face | One-to-Many |
| *Online/Remote* | Electronic | Separated | Sync | Internet | One-to-Many |
| *MOOCs* | Electronic | Separated | Async | Internet | Machine-to-Many |

*Note: The ontology of educational paradigms are not standardized. Digital is often used more generically to mean physical as well as online learning using digital technology. e-Learning is taken synonymously with digital or online learning. Remote learning at times stands to mean Distance learning that may not be online. Traditional or Brick-and-Mortar and commonly use to mean Classroom. Finally, Hybrid as a mix of Classroom and Online.*

**Table 1.**
*Evolution of education systems in India.*

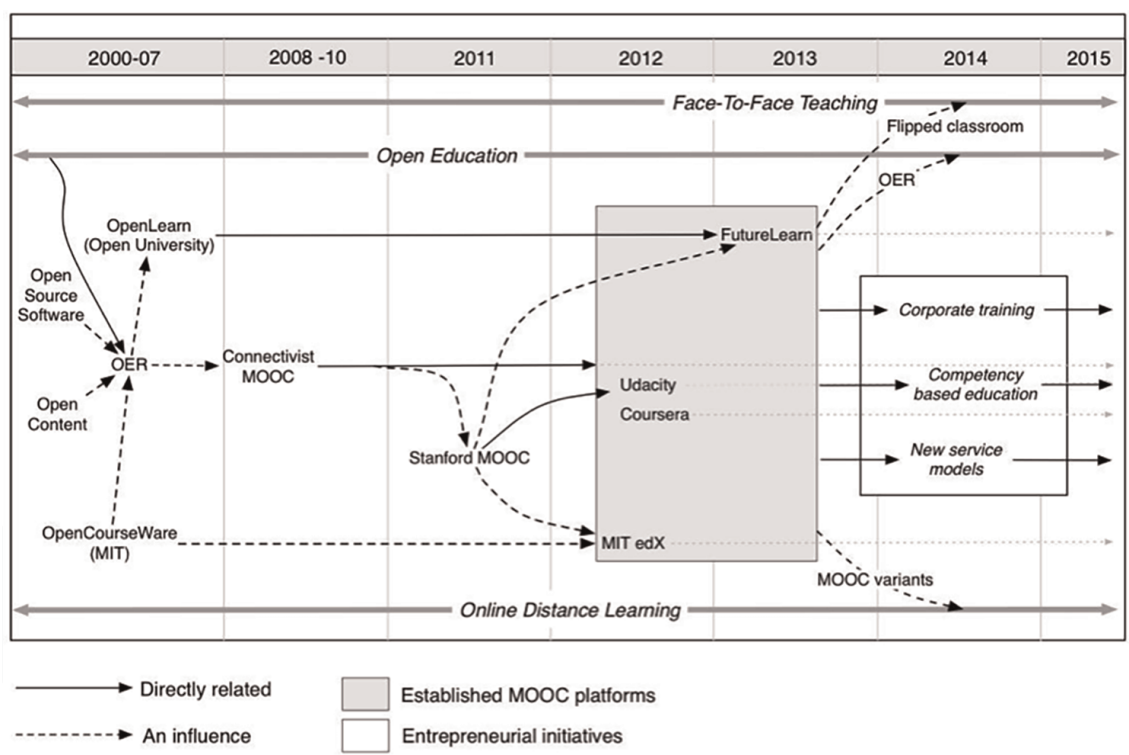

**Figure 1.**
*Milestones in MOOCs and open education.*

(Harvard and MIT), and *Udacity* [18] (Stanford) emerged in 2011–2012 in US. Also, *FutureLearn* [19] (consortium of 12 universities) started in UK by the end of 2012. **Figure 1** shows a simple schematic of this evolution process, MAUT [20] provides a brief history, and Aldahmani et al. [21] investigate the development process of MOOCs to assess their opportunities as well as challenges.

Over the past 10 years 1200+ universities across the world [22] have launched free online courses. Importantly, many countries including *India* (*NPTEL* [23], 2014 &

*SWAYAM* [24], 2016), *China* (*XuetangX*[16] [25], 2013), *Italy* (*EduOpen* [26], 2016), *Israel* (*Campus-il* [27], 2018), *Mexico* (*MéxicoX* [28], 2015), and *Thailand* (*ThaiMOOC* [29], 2017) launched their own MOOCs.

After about a decade of launch, MOOCs world-wide reached an enrollment [22] of 200 million in 2021. As of now, SWAYAM, India's MOOCs, has an enrollment of 30 million and is steadily growing. Consequently, it is important and relevant to understand the various dimensions of MOOCs in India and explore the roadmaps. This chapter attempts to do that.

## 1.2 Motivation

In the last section, we have noted the transformation of education systems in India and the traction of MOOCs in India in the past decade. Recently, India's commitment to MOOCs has reached a new height with *Government of India* (GoI), in its Union Budget 2022–2023, announcing [30] the establishment of *National Digital University* (NDU) to offer exclusively online courses from various public and private universities. Under NDU, courses will start from July 2023 and the students will be able to opt for certificate, diploma, or degree courses with credit mobility through *Academic Bank of Credits* (ABC) [31]. In view of this fast track for use and growth of MOOCs at national scale, it becomes imperative to study the evolution, present status, and future roadmap of MOOCs in India. Hence, we formulate a few research questions the answers to which would be key to the understanding and planning of MOOCs in India.

### 1.2.1 Research questions

1. **RQ1**: How has MOOCs (and its ecosystem) have evolved in India over the decades? What has been its connections to digital and online learning in general? What are the milestones?

2. **RQ2**: What are the innovations in processes and technology that MOOCs in India have created/adopted? Which of these innovations have been key game-changers?

3. **RQ3**: What has been the impact of MOOCs in India in terms of direct and indirect outcome? How is it transforming education in India?

4. **RQ4**: What are the challenges of MOOCs in India? How are these being addressed or can be addressed? Is India missing out on some key opportunities for innovation—especially in technology of MOOCs?

5. **RQ5**: What should be the roadmap for MOOCs in India for the next decade?

It is important to seek answers to these questions and/or accept that more research is needed to answer them as the nation raises various pertinent questions about education in the new age. For example, right before the start of the budget session of 2023–2024, Agarwal [32] asks "*Why Swayam portal's a dampener for Modi govt's online*

---

[16] Primarily fuelled by Tsinghua University and MOE Research Center for Online Education

*learning dream. 3 cr signed up, 4% finished"*—an oft-asked question on poor completion of MOOCs.

Next, we survey the literature to seek answers to these questions and identify the gaps which the current study intends to fill up in this chapter.

## 1.3 Literature survey

We survey two sets of literature to understand the state of the art for the research questions above.

- We study *literature on MOOCs in India* to identify prior work relating to all the five research questions including evolution, innovation, impact, challenges, and roadmap. We survey over a dozen significant articles published during the past 5 years.

- We also study *literature on research on MOOCs in different countries* to contextualize the research questions with respect to the developments across the world. We manage the enormity and complexity of such a study by focusing on about 10 significant review articles on MOOCs research published in the last 3 years.

### 1.3.1 Survey of literature on MOOCs in India

Over the past decade a lot has been written on MOOCs in India. Most of these articles provide *overview of MOOCs projects in India* with reference to infrastructure, investment, diversified needs, quality, web metrics, and challenges. They cover more or less the same ground and discuss similar points barring the temporally changing statistics. Some of the articles in this category, as published in the past 5 years, include Haumin et al. [33], Singh [34], Agarwal et al. [35], Sagar [36], Hooda et al. [37], Mishra [15], EdNet [38], Sharma et al. [39], and Amit et al. [40]. We relate to them in our discussions on questions of impact (Section 4) and challenges (Section 5.2).

In contrast to generic overview, Jaganathan et al. [41], Thakur et al. [42], and Pant et al. [43] *compare SWAYAM with leading MOOCs of the world*—private as well as country-led (including China) and identify a set of challenges for India. We address them in Section 5.2.

Finally, a few recent articles deal with unique questions and/or carry out novel analyses. For example, Varyani et al. [44] use *Power-Interest Grid* as a tool to analyze the problems and concerns of various stakeholders including students, teachers, parents, institutions, content /technology producers, and apex bodies. Singh et al. [45] show high acceptance of MOOCs as an alternative for internship for management students during pandemic and suggest that it may be extend for other domains also during the normal times. Kaicker et al. [46] highlight the *need for Public-Private Partnership for MOOCs* (Section 3.8) and online degrees.

### 1.3.2 Survey of literature on MOOCs research

The research on MOOCs is at least as old as the coinage of the term. Hence, we first take a look at the most cited research articles on MOOCs as compiled by Ledwon et al. [47] through an extensive analysis of citation data on Google Scholar. While these 20 papers have been well-cited and are impactful, they span *2008–2017 only* as any paper would need 3 to 5 years to pile up an impactful citation. Hence, in the fast pace of technology, many of these papers have lost relevance for today's MOOCs. In **Table 2**, we summarize 14 review articles starting with the seminal paper by Liyanagunawardena et al. [48]. Of these, nine papers are from 2019 to 2022 and provide comprehensive current state of the art in MOOCs research through their findings.

| Review reference, Year | # | Period | Findings and remarks |
|---|---|---|---|
| Liyanagunawardena et al. [48] | | 2008–2012 | • Maiden systematic review and quantitatively analysis<br>• Categorization into eight areas of interest |
| Bozkurt et al. [49]; Zawacki-Richter et al. [50] | 362 | 2008–2015 | • Finds four lines of research: (a) the *potential and challenges* of MOOCs, (b) MOOC *platforms*, (c) *learners & content* in MOOCs, and (d) the *quality* of MOOCs and<br>• Bias toward theoretical or conceptual studies; neglect for analyses of actual practice. |
| Al-Rahmi et al. [51] | 32 | 2012–2017 | • Recommends strengthening of—intention to use, interaction, engagement, motivations, & satisfaction. |
| Pradhan [52] | 155 | 2007–2018 | • Only 155 of 4294 articles from India—*it has little contribution* (∼3.6/%); China (∼21.7/%); and USA (∼18.0/%) lead the race<br>• Focus on *issues in computer science* (∼73.5%) and concerns in education & learning sciences (∼43.2%). |
| Zhu et al. [53] | 541 | 2009–2019 | • Observes—quantitative methods, survey & interview for data, and descriptive statistics for analysis.<br>• Focus on *issues of students*, followed by *issues of design*, *context & impact*, and *instructor*.<br>• Attention to *learner retention*, *learning experience*, *social learning*, and *engagement*.<br>• India has ∼10% MOOCs learners [54] of the world, yet *its MOOCs is not studied* here along with US, Canada, Australia, China, Israel, UK, Spain, Sweden, and The Netherlands |
| Jibril et al. [55] | 81 | 2008–2020 | • Stress on the impact of MOOCs on higher education and professional development<br>• Skepticism about MOOCs for school education |
| Meet et al. [56] | 102 | 2013–2020 | • Most researchers based out of United States (22.55%), China (16.67%), and India (11.76%)<br>• Adoption has focus; completion status, instructor-instruction quality, and democratization get less attention<br>• Contradicts the observations by Pradhan [52] who finds marginal research contribution from India. This may be due to the restriction to WoS in [52]. |
| Stracke et al. [57], 2021 | 103 | 2013–2019 | • Use four dimensions—*organizational*, *technological*, *social*, and *pedagogical*—to ascertain the quality of MOOCs.<br>• All but one article focus on only one or two dimensions, often pedagogical dimension.<br>• Assign all 103 articles to the categories of the *Quality Reference Framework for MOOCs* [58], they observe that oft-neglected organizational, technical, and social dimensions are relevant and decisive for the design and quality of MOOCs. |

| Review reference, Year | # | Period | Findings and remarks |
|---|---|---|---|
| Shah et al. [59] | 70 | 2015–2021 | • Dominance of Technology Acceptance Model in adoption<br>• Growth in Asian; slow in developed economies |
| Dalipi et al. [60] | 40 | 2015–2021 | • Students' feedback in MOOCs—*Sentiment analysis* |
| Zhu et al. [61] | 166 | 2011–2021 | • First study on trends and critical issues of *Learning Analytics* (LA) in MOOCs.<br>• LA research is *published in top-tier communication* and used *frequently for research than for practice*.<br>• ∼60% articles *use data science*—learners' log & achievement, ML, network analysis, & visualization<br>• Positive traction of emerging technologies in research, yet lack of actual practice. |
| Cheng et al. [62] | 70 | 2013–2018 | • Research in China by *methods*, *themes*, and *objects*. |
| Despujo et al. [63] | 6320 | 2009–2021 | • Supervised ML to classify to 6 topics—institutional approach, pedagogical approach, evaluation, analytics, participation, and educational resources.<br>• Content analysis by visual network analysis.<br>• Least articles on pedagogical approach & educational resources; most on participation & evaluation.<br>• Analytics and resources most cited; pedagogical and institutional less cited. |

**Table 2.**
*Summary of the reviews of MOOCs research.*

### 1.3.3 Research gaps in literature

Based on the survey of literature, we identify the following gaps:

• Most articles on MOOCs in India are repetitive and provide only generic overview with similar analyses of impacts, challenges, and resolutions.

• No article talks about evolution of MOOCs and its necessary ecosystems in a comprehensive manner.

• There is no attempt to evaluate MOOCs in India with respect to various process innovations.

• Most studies comment on a select set of challenges without interrelating their possible resolutions.

• Research on MOOCs in India as well as from India are both quite inadequate: **Table 2** (Pradhan [52], Zhu et al. [53], and Meet et al. [56]).

• Research on MOOCs is focused more on input (adoption) and much less on output and outcome (completion status, democratization) [56].

- Some of the reviews are contradictory:

- Stracke et al. [57] observe pedagogy as the dominant dimension of research, while Despujo et al. [63] find that pedagogy getting least attention[17].

- While Pradhan [52] finds marginal contribution in research from India and Zhu et al. [53] do not consider India while studying research in nine countries; Meet et al. [56] finds India having the third highest population of MOOCs researchers.

- No review article on applications of emerging technology in MOOCs research like Knowledge Graphs, AI-powered MOOCs, or bots for MOOCs could be found.

## 1.4 Methodology

Given the national (pan-India) scope of the research in this chapter, we decided to use multitude of data sources and research methodologies.

### 1.4.1 Data sources

The major data sources (**Figure 2**) used for the study are summarized below:

1. *Literature on MOOCs*: Over 300 published articles on MOOCs (and related areas in digital/online education) are studied (about 60% of these are listed in the references). The study focused on seminal papers from the early years, extensive topical discovery for the past 5 years, and review papers for the rest.

2. *NPTEL Data (2014–2022)*: Data of MOOCs offered in NPTEL from 2014 to 2022 along with various statistics such as enrolment, exam registration, exam attendance, pass percentage, local chapter engagement, student distinctions. Specific interest is focused on MOOCs impact data and data on various shortcomings and lacunae (gaps).

3. *MOOCs/GoI Sites*: MOOCs in India are significantly covered in GoI and institutional sites that provide a lot of program information along with rich time series data on MOOCs. These include but are not limited to NPTEL, SWAYAM, and IITM BS.

4. *Expert Interviews*: The author had the opportunity to be acquainted with a number of experts (and first-movers) of MOOCs in India. They have been interviewed (as a part of this study or earlier) to provide informative insights into various aspects of MOOCs in India.

5. *MOOCs Hands-on*: The author has been an active MOOCs instructor[18]. In the process, he has gathered a lot of hands-on information on MOOCs and had the opportunity to interact with participants on their feedback.

---

[17] This may be due to skewed sizes of corpus between these two studies.
[18] The author has created three courses in NPTEL [64] that are regularly offered every semester / year. Since 2011, he has been a part of the core team to create and offer IITM's BS in Data Science [65].

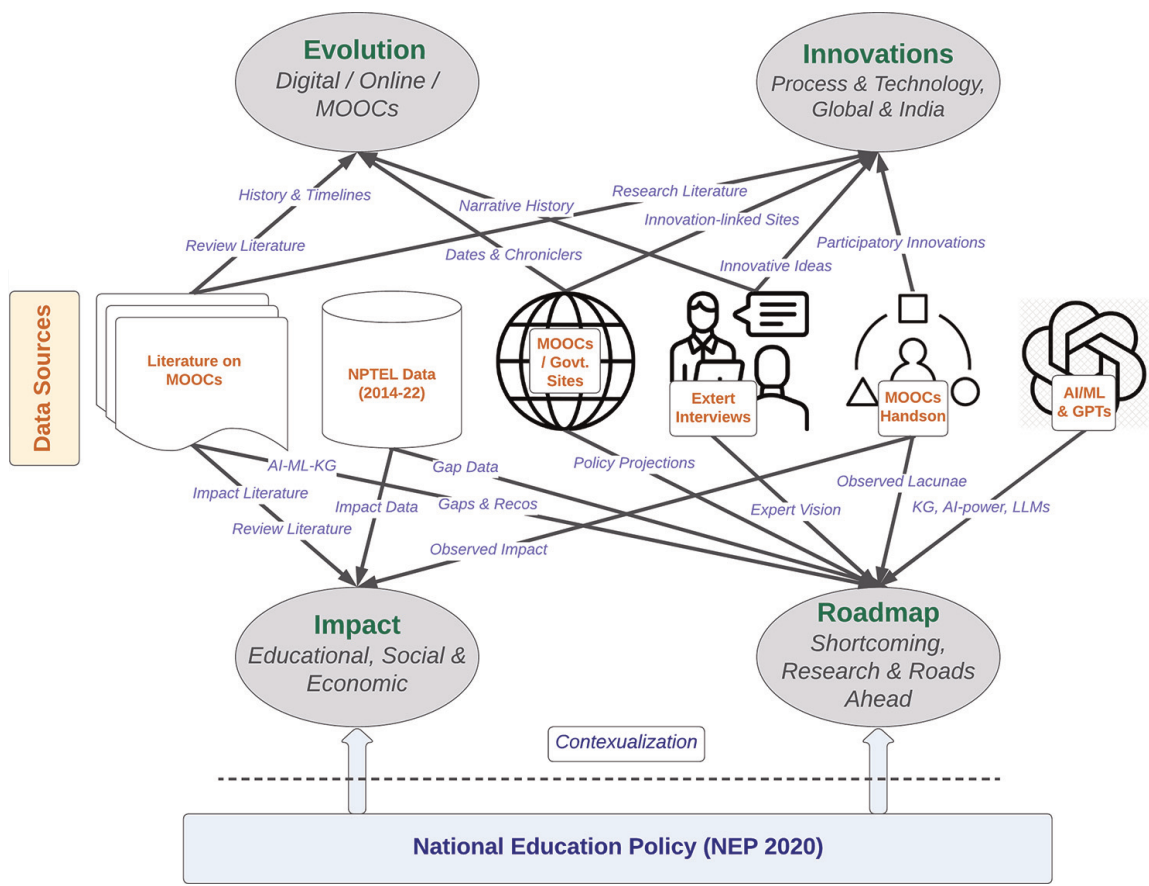

**Figure 2.**
*Methodology framework of the study with major data sources.*

6. *AI[19]/ML[20] & GPTs[21]*: The author has conversed with ChatGPT (and other GPT 3.5 and GPT 4 engines) at length to explore specific AI developments for knowledge graph of MOOCs, AI-power MOOCs, and LLMs in MOOCs. This gives a good perspectives on what research is going on in this area.

*1.4.2 Data analysis*

For each research question, multiple of the sources are used—each with a set of appropriate transformations as shown on the arcs in **Figure 2**.

• **Evolution** *Digital/Online/MOOCs*: To chronicle the evolution of MOOCs in India over decades, we first review articles from MOOCs literature and dates & chroniclers from MOOCs & GoI sites to identify significant events and then arrange them in a timeline with causal relationships. We validate and fill the gaps from interviews with experts. We present a single-page summary[22] in **Table 3** which is subsequently used in the chapter to understand the ecosystem and its causal transformations to assess the innovations, impacts, and challenges of MOOCs in India.

---

[19] Artificial intelligence

[20] Machine learning

[21] Generative Pre-Trained Transformers

[22] A detailed four-page table is available from the author on request.

| MOOCs programs | Year | Open education |
|---|---|---|
| | 1974 | • Working Group on Open University |
| | 1985 | • IGNOU [66] |
| | 1987 | • CET, IITD |
| | 1995 | • CET, IITKGP |
| | 1997 | • ELNet-3 L [67], CET, IITKGP |
| | 1998 | • Visits to CMU |
| | 1999 | • VCTEL<br>• First Virtual Campus of IGNOU |
| • NPTEL [64], IITM (Conceptualized) | 2001 | |
| • NPTEL Phase 1 (Started) | 2003 | • e-Yantra [68], IITB |
| | 2005 | • eGyanKosh, IGNOU |
| • NPTEL Phase 2 | 2006 | |
| • MOOCs coined, cMOOCs started [20] | 2008 | |
| • NPTEL Phase 3 | 2009 | • NMEICT [69]<br>• Virtual Lab [70], IITD |
| | 2010 | • NKN [71]<br>• Baadal [72], IITD<br>• FOSSEE [73]<br>• Spoken Tutorial [74] |
| • Coursera, edX, FutureLearn, Udacity launched | 2012 | • VIDWAN [75] |
| • "Year of the MOOC", says NYTimes [76] | | |
| • XuetangX (China) launched | 2013 | • NROER [77], CIET & NCERT |
| • NPTEL Online Course | 2014 | • NDLI [78], IITKGP |
| • mooKIT [79], IITK | | |
| • IITBombayX [80], IITB | | |
| • IIMBx [81], IIMB | | |
| | 2015 | • e-Pathshala [82], CIET & NCERT |
| • SWAYAM [1] | 2016 | • SWAYAM Prabha [83] |
| | 2017 | • DIKSHA [84]<br>• NMEICT Phase 2 |
| | 2018 | • NDLI [78, 86] Live |
| | 2020 | • NEP 2020 [87]<br>• NISHTHA [88]<br>• NMEICT Phase 3 |
| • Coursera listed in NYSE and edX acquired by 2 U [10] | 2021 | • NITI Aayog partners with Byju's [89] |
| • NPTEL+ | 2022 | • NEAT [90, 91] |
| • Multi-lingual NPTEL | | |
| | 2023 | • NDU [92], PPP for edTech |

**Table 3.**
*Timeline of MOOCs and online education infrastructure in India. Blue items relate to global events.*

- **Innovations** *Process & Technology, Global & India*: Innovations in MOOCs in India have primarily been in the processes of delivery, mentoring, engagement, and empowerment. Hence, to understand the innovations we study the processes in various MOOCs including NPTEL and IITM BS program in depth and identify the introductions of process innovations (with causal necessity and resultant outcomes) at different stages of the timeline. We complete the process using various innovation-guided filters on the data sources as shown in **Figure 2**.

- **Impact** *Educational, Social & Economic*: Impact is primarily assessed from NPTEL Data using various metrics. This is supplemented with literature review (including data from social media like Quora) and the hands-on experience of the author.

- **Roadmap** *Challenges, Research & Roads Ahead*: Challenges are identified from published literature (indirect), MOOCs /GoI sites[23] (indirect), NPTEL data (direct), and hands-on (direct). Availability and projections of emerging technology are extracted from the literature (and technology sites) in consultations with ChatGPT (with factual validations). Finally, experts are consulted on the recommendations of the roadmap.

## 2. Evolution of MOOCs in India

MOOCs in most countries grew bottom-up. Typically, universities created contents and course providers distributed them [93]. For example, we had Coursera and Udacity from Stanford, edX from Harvard and MIT, or FutureLearn and 12 university partners in UK [76]. In contrast, MOOCs in India grew top-down through IGNOU, NMEICT, NPTEL, SWAYAM, and others, where the MoE coordinated the spectrum of initiative from open institutional education and educational content creation, distribution, and certification in associations with HEIs. MoE funded the projects at national level as well. Similar initiatives (Section 1.1.1) have also been taken by *China* [42] (*XuetangX* [25]), *Italy* (*EduOpen* [26]), *Israel* (*Campus-il* [27]), *Mexico* (*MéxicoX* [28]), and *Thailand* (*ThaiMOOC* [29]) where all but XuetangX are free. Notably, no other free MOOCs are as large as SWAYAM.

MOOCs need an ecosystem to exist, grow, and succeed. An ecosystem where *Open Educational Resources* (OER), OER Repositories, *Open and Distance Learning* (ODL), Open Credits and Certification, technology-leveraged processes (like Hybrid mentoring strategies, Industry Associates, Proctored Exams, Internships, and Credit Mobility), and most importantly a open mindset to accept and recognize education beyond the brick-and-mortar, exist and thrive in a harmonious manner. Further, this needs to be supplemented with technology (connectivity, authoring systems, interactions mechanisms, etc.), technical as well as domain expertise, and finances. In India, these happened over half-a-century mostly driven by MoE and partly in an organic manner and we chronicle the significant events of this period in **Table 3**.

---

[23] In terms of plans and policy projections

## 2.1 Open University: Open and distance learning (1974–1985)

While pinning the onset of the process is difficult, we can trace this evolution to 1974 when Mr. Dipanshu Sharma from the Ministry of Education and Social Welfare organized a seminar [94] on *Open University* in collaboration with the Ministry of Information and Broadcasting, the UGC, and the Indian National Commission for Cooperation with UNESCO. On the recommendations of the seminar, the GoI appointed an eight-member working group on the open university under the chairmanship of Prof. G. Parthasarathi, the-then Vice-Chancellor (V-C) of the *Jawaharlal Nehru University* (JNU). The working group recommended (Working Group Report [94], 1974) as follows:

> "... establishing an open university by an act of parliament as early as possible ... the university should have jurisdiction over the entire country so that, any student even in the remotest corner of the country can have access to its instruction and degrees."

On the basis of the recommendations of the working group and several years of deliberations and due diligence, the *Indira Gandhi National Open University Act, 1985 (IGNOU Act 1985)* [13, 14] was passed by both the Houses of Parliament in August 1985 and the *Indira Gandhi National Open University* (IGNOU) came into existence on September 20, 1985. India's march on open education started.

Over 1000 students graduated with diplomas and were convocated in 1989, audio-video courses were the first broadcast by radio and television in 1990, UGC recognized degrees from IGNOU as being equivalent to other universities in 1992, the first virtual campus in India *via* Internet was launched in 1999. Meanwhile, things started moving in other fronts in India.

IGNOU has made a significant mark in the areas of higher education, community education, and continual professional development. It has been networking with reputed public institutions and private enterprises for enhancing the educational opportunities being offered by it. As a world leader in distance education, IGNOU has been conferred with awards of excellence by the *Commonwealth of Learning* (COL), Canada, several times. As of 2020's, IGNOU has [66].

- Nearly 4 million enrolled students

- 21 schools and 67 regional centres

- About 2667 study centres and 29 overseas centres (in 15 countries)

- Approximately 20% of all students enrolled in higher education in India are enrolled with IGNOU.

- Offers 274 academic programs comprising courses at certificate, diploma, and degree levels.

- 810 faculty members and 574 academic staff at the headquarters and regional centres and about 33,212 academic counselors from conventional institutions of higher learning, professional organizations, and industry among others.

## 2.2 Online and digital education (1985-2009)

Soon after IGNOU, India started realizing that open and distance learning cannot scale unless the telecommunications and (then emerging) information & video technology are engaged in it. Also, ODL meant using a pedagogy different from face-to-face classroom teaching[24]. All these needed focused research in what emerged to be known as *Educational Technology* or *EdTech* today.

In 1987, Prof. Anup Roy [86, 95] moved to IIT Delhi from NITTTR, Chandigarh, and setup the country's first *Educational Technology Services Center* (ETSC)[25]. Working with VHS video tapes and miles-long physical wires, ETSC created the first fully interactive remote learning scenarios. Further, Prof. Roy setup *Center for Educational Technology* (CET) at IIT Kharagpur in 1995 and launched *ELNet-3 L: Electronically Networked – Life Long Learning* [67] in 1997. During 1997–1998, CET also connected 15 studios with video labs at Kharagpur, Bhubaneswar, and Raipur[26].

Later in 1998, few directors of IITs and IIMs and some MHRD officials visited *Carnegie Mellon University* (CMU), USA, to learn online learning initiatives by Prof. Paul Goodman who had established an entire University in Mexico functioning completely on the Online Learning model. Impressed by the CMU model, hectic activities started for online education in India and NPTEL was conceptualized in 2001, initially for engineering education. By 2003, NPTEL (Phase 1) led by IIT Madras and collaborated by IITM, IITKGP, IITB, IITD, IITK, IITG, IITRm and IISc started creating video (recording of teaching) as well as web courses (animation with text). It soon created 100 of each type. Further, 700 courses were created by NPTEL (Phase 2) staring 2006. In Phase 3, by 2009, NPTEL moved to HD quality video courses and was scaling up in numbers.

Notably, most of the efforts in this period were focused on creation of quality content and its dissemination to the interested learners. As the time progressed, naturally the quality of recording and editing improved, pedagogy was getting better tuned for online, and the dissemination mechanisms evolved from VHS tapes to VCDs to DVDs/ pen drives to downloadable media on the internet (streaming video was not easy yet[27]).

The initial growth phase for online and digital education was coming to an end and India was getting ready to launch herself at a national scale.

## 2.3 National Mission on education through information and communication technology (NMEICT): Initial period (2009–2014)

*National Mission on Education through Information and Communication Technology* (NMEICT) [69] – the title says it all. The country ventured on a mission mode to impart education by leveraging technology. By the turn of the millennium, India had already been a superpower in ICT serving all across the globe. It decided to use ICT to serve its own educational needs. India committed to the three cardinal principles of **Education for All**: **Access**, **Equity**, and **Quality**, served by:

- providing *connectivity* to all higher institutions,

- providing low cost and affordable *access-cum-computing devices* to students and teachers, and

- providing *high quality e-contents* free of cost to all learners in the country

---

[24] Pedagogy of ODL needs in-depth end-to-end planning, detailed explanations for everything, clear assignments and solutions, and so on

[25] https://etsc.iitd.ac.in/

[26] Internet started in India with *Education & Research Network* (ERNET) project in 1986 and on August 15, 1995, *Videsh Sanchar Nigam Limited* (VSNL) launched public Internet access [96]

[27] The first popular video streaming site, YouTube, was founded in 2005

The Mission encompasses all these elements and has three major components:

1. No talent of the country should be allowed to go waste

2. All the services of content delivery through Sakshat[28] should be free

3. Freely available material on the web should be reused

Following its launch, NMEICT launched several projects and programs in the next 5 years (2009–2014) for the core purposes of

- *Content Creation*: Course contents such as presentations, videos, assignments, and solutions; Contents for learning programming and *Open Source Software* (OSS); Contents for learning Robotics; and Contents for virtually performing experiments

- *Content Delivery*: Online courses maturing to MOOCs; Tutorials for programming; Embedded Systems Kits for Robotics

- *Certification and Credit Mobility*: This happened a few years later in 2016—on full maturity to MOOCs

- *Content Repository and Digital Library*: Open and free repository of digital learning contents created in NMEICT or reused from other sources; Support for various digital library services

- *Support Infrastructure & Activities*: Academic Cloud for high-volume computing; e-Governance for academic campuses; Plagiarism detection for quality research; Unique researcher IDs and database for academic networking; Clubs and competitions for robotics; Local chapters for MOOCs; and so on.

---

[28] SAKSHAT: A One Stop Education Portal, launched on October 30, 2006, was a precursor to NMEICT to facilitate lifelong learning for students, teachers and those in employment or in pursuit of knowledge free of cost to them. https://www.education.gov.in/technology-enabled-learning-1

| NMEICT [69] Platforms for Digital Learning | | | |
|---|---|---|---|
| **Project** | **Platform** | **Level** | **Purpose** |
| SWAYAM [83] & (NPTEL) [64] | MOOCs | School, UG, PG | Contents, Delivery & Certification |
| SWAYAM Prabha [84] | 34 DTH Channels | School, UG, PG | Contents & Delivery |
| Spoken Tutorial [74] | Learn Programming | All learners[a] | Contents |
| Fossee [73] | Learn OSS | UG, PG | Contents |
| e-Yantra [68] | Learn Robotics | UG, PG | Contents Activity[b] |
| Virtual Labs [70] | Scientific Experiments | UG, PG | Contents |
| NISHTHA[c] [88] | Teachers' Training | School | Contents & Delivery |
| NDLI[d] [78] | Open & Free Contents | All learners | Library |
| e-ShodhSindhu [97] | e-Journals Subscriptions | UG, PG | Library |
| eGyanKosh [98] | ODL[e] Institutions | UG | Library |
| e-PG Pathshala [99] | UGC PG | PG | Delivery |
| Shodh Shudhhi [100] | Plagiarism Detection | UG, PG | Infrastructure |
| Samarth [101] | Campus Governance | UG, PG | Infrastructure |
| Baadal [72] | Free Academic Cloud | UG, PG | Infrastructure |
| VIDWAN[f] [75] | Experts Database | UG, PG | Infrastructure |

[a]School, UG, PG, Teachers, Life-long learners, Working professionals.[b]Robotics Clubs, and Regional and national competitions. Delivering indigenous kits for robotics.[c]National Initiative for School Heads' and Teachers' Holistic Advancement.[d]National Digital Library of India.[e]Open and Distance Learning.[f]VIDWAN means learned, knowledgeable.

**Table 4.**
*Digital platforms in India (NMEICT, MoE).*

These have been summarized in **Table 4** and put in the timeline in **Table 3**. In addition to the NMEICT fold of projects, *National Knowledge Network* (NKN) [71] was established in 2010 for high-speed backbone connectivity between 1500+ institutions to enable knowledge and information sharing toward ushering in a knowledge revolution in the country.

Thus, during 2009–2014, India got ready with a strong ecosystem to go the MOOCs way at the national scale [102]. As of 2023, the footfall to different major projects as reported from NMEICT [69] site are:

- SWAYAM: 9,20,027 (9.2 lakh ≡ 920 thousand)

- SWAYAM PRABHA: 1,25,07,512 (1.25 cr ≡ 12 M+)

- NDLI: 3,18,46,733 (3.2 cr ≡ 32 M)

- SPOKEN TUTORIALS: 1,79,42,101 (1.8 cr ≡ 18 M)

- ICT INITIATIVES: 3,08,436 (3.1 lakh ≡ 310 thousand)

## 2.4 Massive open online courses: NPTEL-SWAYAM (2014–2023)

Though NPTEL has been offering online courses since 2006 in multiple forms, these have not been as complete as normal courses to be used for curricular requirements. Students loved NPTEL courses for *Quality*—the quality of the teachers, the quality of teaching, and the quality of contents. Yet, they could use these courses only for better understanding and leaning, but were not being able to use them for their curricular requirements. And many, naturally, wanted to.

### 2.4.1 NPTEL

So NPTEL decided to go the full-fledged MOOCs way and started *NPTEL Online Certification* [64] (NPTEL-NOC) in 2014. About 600 video and web-based courses came to the fold with a well-developed four-quadrant pedagogy (**Table 5**). A course in NPTEL got structured as follows:

- A course has a duration of 4, 8, or 12 weeks.

- Every week the student gets to download 5 modules for the week.

- Each module is discussed in a 30-minute video with presentation and video transcription files.

- Every week has an assignment which the student has to complete and upload within a stipulated time. It is evaluated and the performance is shared with the student. The model solution is also shared.

- The students also connect to a discussion forum where doubts are discussed and clarified.

- An interactive live session is conducted with the instructor once every 4 weeks.

- At the end of the course, a proctored examination is conducted physically in multiple cities where the student can take the examination showing an official ID.

| *Quadrant-I is e-Tutorial*: <br> • Video and Audio Content in an organized form <br> • Animation, Simulations, Video demonstrations, Virtual Labs, etc. | *Quadrant-II is e-Content*: <br> • PDF, Text, e-Books <br> • Illustrations, video demonstrations, documents <br> • Interactive simulations wherever required. |
|---|---|
| *Quadrant-III is Web Resources*: <br> • Related Links, Wikipedia Development of Course <br> • Open source Content, OER <br> • Case Studies, books including e-books, research papers, & journals | *Quadrant-IV is Self-Assessment*: <br> • Problems/Solutions: MCQ, MSQ, SA, LA, Matching <br> • Assignments and solutions <br> • Live Sessions <br> • Discussion forum and FAQs |

**Table 5.**
*Four quadrant pedagogy of NPTEL [64] and SWAYAM [83].*

- Finally, the student gets the grade based on the performances in the assignments and the final examination. Passing students get a certificate from NPTEL. There are further distinctions to earn for good performances.

- The course is free to attend (including assignments), while the final examination has a nominal fee.

As of January 2023, NPTEL bears the following stats [103]:

- 1.6 Billion+ views

- 4.4 M+ YouTube subscribers

- 2500+ unique courses available for self study

- 2.07 Cr. + enrollments

- 23.8 Lakh+ exam registrations

- 5322+ LC colleges

- 4707 MOOCs completed

- 70+ Industry associates

- 60,900+ hours of videos English subtitles

Riding on the acceptability of NPTEL courses, NPTEL-NOC soon proved to be very popular. Thousands of students started learning from them. This led to two questions:

- Should MOOCs be limited to NPTEL, that is, primarily in engineering and science?

- How can a student transfer the earned credit from NPTEL to college/university transcript?

In response to the first question, SWAYAM [83] was born in 2016, and a while later a new regulation was passed in response to the second question.

## 2.4.2 SWAYAM

SWAYAM [83] is designed to cover all aspects and disciplines of education. To ensure that best quality content is produced and delivered, nine *National Coordinators* (NC) have been appointed with respective focus areas:

1. AICTE (All India Council for Technical Education) for Self-paced and International courses

2. NPTEL for Engineering

3. UGC for Non-technical post-graduation education

4. CEC (Consortium for Educational Communication) for Under-graduate education

5. NCERT (National Council of Educational Research and Training) for School education

6. NIOS (National Institute of Open Schooling) for School education

7. IGNOU or Out-of-school students

8. IIMB (Indian Institute of Management, Bangalore) for Management studies

9. NITTTR (National Institute of Technical Teachers Training and Research) for Teacher Training program

The NCs have the mandate to drive SWAYAM within its specified guidelines [104]. While SWAYAM follows the same four-quadrant pedagogy (**Table 5**) like NPTEL, depending on the diversity of levels of education, requirements, and discipline, it follows multiple different course delivery and evaluation structures. In 2019, Singh [34] studied the working of the SWAYAM platform in respect of navigation, content management, interactions & communications, and assignment management from the responses of the faculty who have been developing MOOCs. Many of its recommendations got adopted over time.

As of October 2022, the major stats of SWAYAM have been as follows:

• 9.5 M+ Total registered Users

• 9 National coordinators: Operated by AICTE, NPTEL, UGC, CEC, NCERT, NIOS, IGNOU, and NITTTR

• 8000+ Total number of Courses hosted so far

• 2 M+ Exam registrations over 7 years

• 2.6 M+ July 2022 Semester enrollments

*2.4.3 Credit mobility of MOOCs*

UGC soon followed up with the regulation *Credit Framework for Online Learning Courses through SWAYAM* [105, 106] in 2016 and allowed (limited) credit mobility [107]. Credit Mobility has given a real boost to MOOCs as a strong supplementary arm of education is India which can play a deep curricular role. We discuss the details in Section 3.1.

*2.4.4 Institutional MOOCs*

While MoE led the country with the umbrella of NMEICT for MOOCs and related online education initiatives, there have been other efforts by institutions on

| Other platforms for digital learning | | | |
|---|---|---|---|
| Project | Platform | Level | Purpose |
| mooKIT [79] | MOOCs[a] | All learners[b] | Contents, Delivery & Certification |
| IIT Bx [80] | MOOCs[c] | All learners | –do– |
| IIM Bx [81] | MOOCs[d] | PG | –do– |
| NROER[e] [77] | Multi-format Books & Flipbooks | School | Contents |
| e-Pathshala[f] [82] | Learning Contents | School | Contents activity |
| NKN [71] | Pan-India Network | | Infrastructure[g] |
| DIKSHA [85] | Knowledge Sharing | School | Infrastructure |
| NEAT [90] | EdTech Products | | Infrastructure[h] |
| Electronics & ICT Academy [108] | Financial Assistance | | Infrastructure[i] |
| Shagun [109] | Integrated Information | School | Infrastructure |
| NDU [92] | Virtual University | UG & PG | Infrastructure[j] |

*Many other MOOCs or other online platforms are studied in Refs. [33, 41].*
[a]*Connect with IIT Kanpur for 150+ courses in India and abroad having 2,00,000+ learners from 160+ countries.*[b]*School, UG, PG, Teachers, Life-long learners, Working professionals.*[c]*Offers Hybrid MOOCs with flipped classrooms, online lectures, and live sessions in multiple flavors—EduMOOCs (as extension courses), SkillMOOCs (vocational training), TeachMOOCs (for teachers), LifeMOOCs (for working professionals).*[d]*Offers MOOCs on SWAYAM, edX and IIMBx (open edX).*[e]*Developed by Coimbatore Institute of Engineering and Technology (CIET) and NCERT.*[f]*Developed by CIET & NCERT. Besides contents, students participate in exhibitions, contests, workshops; teachers /educators participate in research, and parents can learn to nurture creative talent.*[g]*A high-speed backbone connectivity for knowledge and information sharing between 1500+ institutions.*[h]*A PPP model between the GoI (through AICTE) and the EdTech companies of India. Selected companies showcase products on a national portal for the learners to purchase.*[i]*MeitY (through IITG) provides Financial Assistance to Set up Electronics and ICT Academies.*[j]*First digital university of India—will operate as a hub for flexible online education.*

**Table 6.**
*Digital platforms in India (besides NMEICT).*

specialized MOOCs. Year 2014 saw the emergence of three pioneering programs—
*IIMBx* [81] from IIM Bangalore, *IITBombayX* [80] from IIT Bombay, and *mooKIT* [79]
from IIT Kanpur. These are summarized in **Table 6**.

IIM Bangalore partnered with edX, a not-for-profit (at that time) online initiative
of Harvard and MIT, to create IIMBx [81] platform for MOOCs in all areas of man-
agement—data and insights, economics, finance, marketing, people management,
operations, and strategy. Apart from edX, the program offers courses on its own
platform and SWAYAM for every kind of learner, ranging from first-generation
entrepreneurs to college educators looking to teach better.

IIT Bombay, on the other hand, invested in IITBombayX Hybrid MOOCs Platform
[80]. Its novelty hovered around its hybrid nature in the forms of flipped classrooms,
online lectures, and live sessions. However, what stands out in this initiative is the
user-centric flavor where special MOOCs are designed and offered for Extension
courses (*EduMOOCs*), Vocational training (*SkillMOOCs*), Training of teachers
(*TeachMOOCs*), and Courses for life-long learner and working professionals
(*LifeMOOCs*).

Last but not the least, IIT Kanpur offered mooKIT [79] with 150+ courses in India and abroad for 2,00,000+ learners from 160+ countries.

MOOCs started booming from the past 10 years.

### 2.4.5 Learning support systems

SWAYAM, NPTEL, and institutional MOOCs aside, there have been other efforts by ministries and departments of GoI to propel advances in this area (See **Table 6**). Notable among these are *National Repository of Open Educational Resources* (NROER) [77] in 2013 (with Video, Image, Audio, Document, and Interactive for schools with NCERT books in Flip book format) and e-Pathshala [82] in 2015 (with resources in English, Hindi and Urdu for teachers, students, parents, researchers, and educators on the web and the mobile). In 2017, *Digital Infrastructure for Knowledge SHAring* (DIKSHA) [85], a knowledge sharing platform for schools known as *One Nation, One Digital Platform* under PM's *eVidya* initiative of *Atma Nirbhar Bharat*, started providing DIKSHA in NCERT Textbooks 36 Indian languages for learners and teachers of India.

## 2.5 National Education Policy: NEP 2020

The story of MOOCs in India reached out for a climax beyond MOOCs as India announced its much-awaited *National Education Policy* [87] in 2020 (NEP 2020) replacing NEP 1986 [110]. NEP 2020 outlines the vision of new education system of India. It is a comprehensive framework for elementary to higher education as well as vocational training in both rural and urban India. It targets to transform India's education system by 2030 through *National Educational Technology Forum* [111] (NETF)—a platform for the free exchange of ideas on the use of technology to enhance learning, assessment, planning, administration. While the general response to NEP has been quite positive, there have been legitimate voices of concern [112] especially regarding its implementation.

NEP 2020 is structured in terms of multiple clauses in different sections covering different aspects of the education system. Highlights from clauses focusing on online digital education including MOOCs are given below:

- *Online Courses*: apps, online courses, TV channels, online books, and ICT-equipped libraries (Cl 21.9, p. 52), Life-long learning (Cl 21.10, p. 52)

- *Digital Learning Contents*: Digital libraries for contents and textbooks (Cl 24.4 (d), p. 59). Coursework, Games & Simulations, AR & VR. Textbooks (Cl 21.9, p. 52) (Cl. 4.32, p. 17)

- *Indian Languages*: Education in all languages in India (Cl 22.19, p. 56)

- *Pedagogy & Teaching–Learning Process (TLP)*: Digital Pedagogy & enriched TLP with online resources, Technological interventions in TLP (Cl 23.5, p. 57). Teacher Education. Educational Software and Contents (Cl. 23.6, p. 57)

- *Smart Infrastructure*: Digital classrooms with appropriate technology (Cl. 4.46, p. 20)

- *Flexible Processes*: Flexible Curriculum & entry-exit, Assessment & Certification, and Flexible entry-exit

- *Equitable and Inclusive Education*: Learning for All and in HE (Cl. 6, p. 24)

## 2.6 National Digital University

The establishment of *National Digital University* (NDU), as envisioned under NEP 2020 [87], has been announced [30] in the Union Budget 2022–2023. NDU [113] will be setup by *Department of Higher Education*, MoE, in consultation with UGC, AICTE, and other stakeholders. On this, Mr. Mayank Kumar, Chairman of the *India Edtech Consortium* (IEC)[29] observes [92, 114]:

> *"NDU could be India's next UPI[30] moment, making quality education more accessible and flexible."*

The salient features of NDU [115] include:

- NDU will function in a *hub-and-spoke model* [116] where SWAYAM and Samarth [101] will be the hub and universities and higher educational institutions will collaborate as a network of spokes.

- *All government-funded institutions* such as IITs, NITs, IIITs, and central universities (and a few private universities) would be a part of the network.

- NDU will allow students to *register for multiple courses at different institutions*[31], accumulate credits from them in *Academic Bank of Credits* (ABC) [31], and redeem these to get a certificate, diploma, or a degree.

- Students will be *awarded degrees* by the university where they earned 50% of the credits. If the credits are thinly spread, NDU will award the degree.

- There will be *no quota of seats*—any number of students can join, based only on the passage of the qualifying exam.

Union Budget 2023–2024 reiterated the plans for NDU [119] and announced [114] that courses will be offered from July 2023.

*In summary, we have discussed the evolution of MOOCs, its ecosystems, and its synergy with digital and online learning over decades. We have also chronicled the milestones in a timeline while addressing* **RQ1** *(Section 1.2.1)*.

---

[29] https://www.indiaedtech.in/

[30] India's Universal Payments Interface (UPI) has transformed the digital payments system in the country and is extremely popular.

[31] In April 2022, the UGC approved simultaneous dual degrees, in both physical and online modes [117]. Also, there are 500+ UGC-approved online degrees from India's top universities [118]. With NDU as hub for everything online, it is not clear how this will be integrated.

## 3. Innovations in and extensions of MOOCs in India

MOOCs have made significant impact in India through SWAYAM in general and NPTEL in particular. The early success of NPTEL (focusing on Engineering and Science) in the first decade of the millennium paved the way for SWAYAM with MOOCs in all disciplines, at all levels, and in multiple forms of education. Based on the learning from NPTEL MOOCs, SWAYAM/NPTEL continued to innovate in various programs [103, 120] to enrich the ecosystem on MOOCs to enhance its traction and effectiveness. We highlight a few key innovations that happened on the way.

### 3.1 Credit mobility

To understand the credit mobility, let us see how credit is computed.

#### 3.1.1 Credit computation

Every 4 weeks' course entails 14 hours of academic work for the student:

- *Videos*: 30 minutes/video $\times$ 5 video/week $\times$ 4 weeks = 10 hours of video-based self-study

- *Assignments*: 1 hour/assignment $\times$ 1 assignment/week $\times$ 4 weeks = 4 hours of assignment workout

According to UGC norms [120], 14 hours of academic work is worth for 1 credit. Similarly, 8 or 12 weeks' courses are worth for 2 or 3 credits, respectively. In addition, if a college wants to count the discussion forum, extra material, and more time for videos/assignments, 1 extra credit can be assigned to the courses.

#### 3.1.2 Credit mobility and management

UGC allowed (limited) credit mobility [105, 106] in 2016 (Section 2.4.3). Under this an institution can allow up to 20% of the total courses in a semester through SWAYAM. Any regular/part-time student of any educational institution in India [121] can avail of the credit transfer. Credit mobility soon became popular and in 2021, the limit on credit mobility per semester [122] was increased to 40%.

Further, to facilitate easy management and mobility of credits, UGC has setup [123] the *Academic Bank of Credits* [31] in 2021. Institutions register to ABC and deposit credits earned by the students' to their respective accounts. This helps to maintain the integrity, authenticity, and confidentiality of student credits, easy credit transfer in digital mode, and fast credit recognition.

Credit Mobility has given a real boost to MOOCs as a strong supplementary arm of education is India, which can play a deep curricular role. Recently, Singh and Kakkar [124, 125] have analyzed the impact of credit mobility to reveal that there has been a considerable increase in student enrollment (due to the mobility) but with extremely low certifications (We discuss this aspect in Section 5.2).

**3.2 Certification**

MOOCs usually have certificates for courses successfully completed. However, India has been desirous of creating certificates that will be *credible and acceptable to all academic institutions* and also define the *scope of skills the student is certified for*.

*3.2.1 Credible certification*

To add credibility to the SWAYAM certificates, NPTEL-NOC had introduced a *physically proctored examination* conducted in multiple cities where the student can *take the examination showing an official ID*. While weekly assignments, their solutions, and evaluations comply with the continuous learning principle of SWAYAM (MOOCs), award of final grade adequately weighted with the performance of the proctored examination mean credibility of the certification. Three levels of certifications have been created over a decade:

- *Course-level certificate* (as in NPTEL-NOC) certifies skills in the course

- *Domain-level certificate* (discussed below) certifies skills in a domain of specialization comprising multiple courses.

- *Degree-level certificate* (discussed under Online Graduation) certifies comprehensive skills in a discipline—at par with what is obtainable from brick-and-mortar curricula.

*3.2.2 NPTEL domain certification*

NPTEL has been giving out course-level certificates since 2014 and there have been a lot of students doing multiple courses from NPTEL, which are not always connected. Hence, a need has been felt to link courses together for the foundations (*core*) and to define baskets of electives for further specialization. NPTEL Domain Certification has been created for this purpose. In this, NPTEL has grouped courses across 12 disciplines to help learners specialize in 51 domains. Completing a domain helps to gain expertise in a specific area [126]:

- To gain expertise/foundations in an area of interest

- To gain mastery to pursue Higher Education

- To become more employable for jobs in the opted area

Every domain comprises Core course and Elective courses (to choose from a list). The learner has to complete the courses with at least 55% score in each and at least 60% in aggregate to get a domain certification.

**3.3 Online graduation**

Building up on online certification, which is for specific courses and domain certification, which is for a basket of courses from a domain, IIT Madras launched the world's first 4-year Bachelor of Science (BS) Degree in Data Science and Applications [65].

This gives the students an opportunity to work toward an undergraduate degree/diploma from an IIT regardless of her/his age or location, and with a wide range of academic backgrounds.

This was launched in January 2021 as a BSc Degree in Programming and Data Science from IIT Madras. Keeping with the structure suggested by NEP 2020 [87] it is designed as a multi-entry and multi-exit Online Degree program where a student can earn certification at Foundational, Diploma, and Degree levels. Subsequently, IIT Madras launched 4-year BS in Data Science and Applications [65] and Diploma in Programming & Diploma in Data Science [127].

### 3.4 Partnering and networking

From the early days NPTEL, and later SWAYAM, have been built on a strongly participatory model between various institutions. These partnerships and networks have been designed for course creation, delivery, and certification. On way to NPTEL-NOC, NPTEL realized that the success of MOOCs as a large-scale model for affordable education needs the engagement of students and the colleges that the students attend. Hence, the concept of *Local Chapters* (LC), at the colleges partnering NPTEL, was born.

As the industry engages the skilled man-power, SWAYAM/NPTEL needs to have close collaboration with it. The *Industry Associate Program* thus created by NPTEL, helps employment and course monitoring for effective skilling.

### 3.4.1 NPTEL local chapters

Following the launch of NPTEL-NOC [64], NPTEL started setting up LCs in colleges to encourage more students to participate in it. Every LC is headed by a faculty member of the college, who acts as the *Single Point of Contact* (SPoC) between the college and NPTEL. The SPoC disseminates information about programs among the students, identifies suitable mentors for courses, ensures that the students are active in courses, clarify their doubts, and so on.

> *Local Chapter has been the most effective outreach for NPTEL. Over the past five years (2017-22), about 70 to 80% of total NPTEL registrations have taken place through the LCs [128]. States having more LCs have shown more engagements with NPTEL.*

### 3.4.2 NPTEL industry associate (NIA) program

NPTEL partners with industry to bridge the gap between the academics and the industry and to create courses along with the industry to cross-skill and upskill the existing workforce. CSR initiatives are also welcomed as part of this association.
A total of 79 NIAs [103, 120, 129] have signed up and facilitated in one or more areas:

- *Recruitment/Internship*: NIAs are regular recruiters of NPTEL learners and are offering internships

- *Co-offer Courses*: NIAs have collaborated in offering 170+ live sessions in NPTEL Special Lecture Series on Latest technologies, Skills or competencies for the industry, and Career opportunities

- *CSR Support*: Through CSR, NIAs are supporting waivers for examination fees (∼1,20,000+ students benefited)

- *Soft/Digital Skilling*: NIAs support to conduct soft skills training and digital skilling courses to learners in local chapters.

- *Upskill/Reskill Employees*: Courses for working professionals to reposition in career.

## 3.5 Student connect and recognition

Students are the purpose of MOOCs. So keeping them encouraged, motivated, and engaged are critical for effective MOOCs at scale. NPTEL started this through various honor programs for students and also created opportunity for internship at academia as well as industry.

### 3.5.1 NPTEL star

To motivate learners to excel, NPTEL has introduced multiple honors[32] under the *NPTEL Star* program. Starting the Jul-Dec, 2019 session, over 10,000 certificates have been awarded to stars in various categories in the past 3 years. This has proved to be quite motivational from the learners [130].

### 3.5.2 NPTEL internship

From 2018 summer, NPTEL has started offering internships to NOC examination toppers with the respective course instructors to provide an opportunity to toppers to gain rich research experiences. The internship is offered for 4, 6, or 8 weeks in summer and winter each year with a stipend of Rs. 5000/= for 4 weeks of internship. A total of over 200 students [103, 120, 131] took internship in summer and winter sessions of 2021 and 2022.

## 3.6 Extensional courses

NPTEL has also launched extensional courses for students (GATE), teachers (FDP), and industry professionals (NPTEL+) appropriately reusing the IP already created for NPTEL courses. It has high return on low investment.

### 3.6.1 NPTEL GATE project

To provide students with an integrated platform to prepare for the *Graduate Aptitude Test in Engineering* (GATE)[33], NPTEL GATE Project was initiated with CSR

---

[32] NPTEL Domain Scholars, Superstars, Evangelists, Motivated Learners, Enthusiasts, Discipline Stars, and Believers

[33] GATE is an examination conducted in India that primarily tests the comprehensive understanding of various undergraduate subjects in engineering and science for admission into the Masters Program and Job in Public Sector Companies. GATE is a *Computer-Based Test* (CBT) organized by IISc Bangalore, IIT Bombay, IIT Delhi, IIT Guwahati, IIT Kanpur, IIT Kharagpur, IIT Madras, and IIT Roorkee, on behalf of the

support from Amadeus Labs, Bengaluru, India. This project supports preparations for GATE in multiple ways including in-video contents for [132]:

- *Video solutions to previous GATE questions*: The video solutions are designed to be self-contained—starting from a quick recap of the basic concepts for solving the problem, followed by a detailed solution using fundamental concepts, and ending with smart tricks, if any, to solve the problems.

- *Video explanation of Subject Concepts*: For easy access, all the topics of GATE syllabus are mapped and linked to the most relevant NPTEL lecture/s. These lectures, in turn, are mapped to various topics of the GATE Syllabus.

With solutions for 9 departments, 4191 videos, ∼420 hours of content, and ∼ 800 live mentoring sessions in 2 phases, 6500+ students have benefited [103].

### 3.6.2 NPTEL courses as FDP

In July 2018, NPTEL signed an MoU with AICTE for advanced NPTEL online certification courses [133, 134] approved for *Faculty Development Program* (FDP) by AICTE. About 4, 8, or 12 weeks courses are offered as $\frac{1}{2}$, full or $1\frac{1}{2}$ FDP of 1 week. 2400 + FDP courses [103] have been offered during January 2021 to April 2023 with 66,000 + faculty applying for FDP during January 2021 to July 2022.

### 3.6.3 NPTEL+: Anyone, anywhere, anytime

NPTEL deviated from *free* paradigm of MOOCs and designed paid e-learning courses for working professional as NPTEL+ portal to expand the variety of offerings and for learner upskill. Three types of training programs are available [135]:

1. *NPTEL courses in self-paced mode*: These are self-paced courses where learners may progress through the course and complete assignments at their own pace. Learners may also choose to write a remote proctored online exam from the comfort of their homes and earn a certificate.

2. *Short-term training programs from the IITs/IISc*: Short-term training programs which might involve fully live lectures coupled with hands-on training or a blended mode of learning (recorded videos+live lectures) are planned.

3. *Other programs*: These are targeted toward specialized courses in an emerging technology or complementing the existing NPTEL courses with dedicated hands-on content to equip the learners to be industry ready.

With 200+ courses and 550+ professionals already qualified [103], NPTEL+ is leveraging the IPs of NPTEL MOOCs to add new value for the online courses. Extensions of these are offered for in-person lab certification through workshops:

---

National Coordination Board – GATE, Department of Higher Education, MoE. GATE 2023 is coordinated by IIT Kanpur.

- *NPTEL lab workshops*: Typically offered during summer and winter, these in-person courses motivate students and faculty to get hands-on experience.

- *NPTEL+ workshops*: These live and interactive workshops are offered through the year in online or in-person mode for half-a-day to a week. They cater to a wider audience including students, faculties, and professionals.

A total of 24 NPTEL+ course have been offered during August 2022 to December 2022 where 4000+ learners have participated [103].

## 3.7 Multi-lingual MOOCs

In India there are 22 languages [136], referred to as *scheduled languages*, and given recognition, status, and official encouragement. About 96.71% of the population in the country has one of the 22 scheduled languages as their mother tongue. Several students undergo their schooling in their regional language and may face challenges transitioning to English for technical education.

The MOOCs courses, however, are primarily in English. In the early years of development (2000–2010), this was not considered a bottleneck as MOOCs meant courses in engineering, science, and management where the teaching in classrooms are also only in English medium. However, with the widening of education with online and digital options, decision of using MOOCs at all levels (including school education, adult education, and life-long learning), and broadening of outlook in education across the Nation, it became imperative that skills in English cannot be a prerequisite for education through MOOCs. Hence, multi-lingual MOOCs have become a necessity for India. Given the diversity of disciplines and levels, and variety of languages, multilinguality is a major challenge. Yet, NPTEL had an early start which SWAYAM too is adopting.

### 3.7.1 NPTEL translation

NPTEL has initiated translation of course contents into *11 different languages*— Assamese, Bengali, Gujarati, Hindi, Kannada, Malayalam, Marathi, Odia, Punjabi, Tamil, and Telugu. Translations are available as pdf transcripts, e-books, subtitled videos, scrolling text-on-videos, and audio files. So far, NPTEL has received 1500+ requests to provide the translation of course contents. With the help of LCs[34], it has translated ∼300 courses in 11 languages having 40,000+ lectures for 10,000+ hours [103, 137]. These are freely accessible and downloadable.

Through this intuitive, SWAYAM/NPTEL has been supporting the need for *education in all languages in India* as envisioned in NEP 2020 [87] (Cl 22.19, p. 56).

## 3.8 Public-private partnership (PPP)

With the fast increasing need for skilling India's workforce, huge resources are continually needed. A part of that is addressed by the structure and efficiency of MOOCs. Yet it is beyond government's resources to satisfy all the need of skilling; and participation of the private sector is a dire necessity. Kaicker et al. [46] have drawn

---

[34] Over 3500 translators and 200+ curators from nearly 1500 LCs have been contributing

parallels between SWAYAM and edTech start-ups in India that offer online degrees in collaboration with difference universities to highlight the complementarity between private and public sectors in India in this space.

Keeping the above in view, under NEP 2020, EdTech companies and startups are provided with necessary guidelines and impetus to develop learning management systems, ERP softwares, assessment platforms, online labs, etc., for schools and universities. NETF, an autonomous body mandated by NEP 2020, is created to facilitate exchange of ideas on technology usage to improve learning [138]. Government is also experimenting with different mechanisms to make the PPP successful and effective. We highlight two recent initiatives.

### 3.8.1 PPP of NEAT

MoE has announced a *National Educational Alliance for Technology* [90] (NEAT) as a *Public-Private Partnership Model* between the GoI (represented by AICTE) and the EdTech companies of India. The aim of NEAT is to bring the best products in educational pedagogy on a single platform for the convenience of learners. Technology Products using AI for customized learning or e-content in niche areas having highly employable skills would be identified for showcasing on the portal. In September 2021, *NITI Aayog*[35] partnered with Byju's[36] to provide free access to its tech-driven learning programs to engineering aspirants from 112 districts [89].

### 3.8.2 Electronics and ICT academy

In November 2014, MeitY[37] launched *Scheme of Financial Assistance for setting up of Electronics and ICT Academies*[38] for faculty/mentor development/upgradation in fast evolving areas in electronics and information technology. Under the Scheme, seven Electronics and ICT academies have been set up at seven premier academic institutions—NIT Warangal (Telangana), IIITDM Jabalpur (Madhya Pradesh), IIT Guwahati [108] (Assam), NIT Patna (Bihar), IIT Kanpur (Uttar Pradesh), IIT Roorkee (Uttarakhand), and MNIT Jaipur (Rajasthan). The scheme targets to train 92,800 faculties.

> To address **RQ2** *(Section 1.2.1), we have discussed various process innovations in MOOCs in India either through the initiatives of the government or the MOOCs providers like SWAYAM/NPTEL. Some of these like credit mobility, domain certification, online graduation, local chapters, and multilingual support are proving to be game-changers and need to continue in the future roadmap.*

---

[35] The NITI Aayog (*Policy Commission* or National Institution for Transforming India: https://www.niti.gov.in/) serves as the apex public policy think tank of the GoI

[36] Byju's (https://byjus.com/) is an Indian multinational educational technology company, headquartered in Bangalore, India. It was founded in 2011 and as of March 2022, it is valued at US$22 billion and the company claims to have over 115 million registered students.

[37] Ministry of Electronics and Information Technology: https://www.meity.gov.in/

[38] https://www.meity.gov.in/scheme-financial-assistance-setting-electronics-and-ict-academies

## 4. Impact of MOOCs in India

NPTEL completes 20 years in 2023. This by itself is a strong corroboration of the impact it has been making in MOOCs-based education. Over the past decade, a lot has been written about the impact of SWAYAM/NPTEL by several researchers including Chauhan et al. [139], Jaganathan et al. [41], Haumin et al. [33], Varyani et al. [44], Pant et al. [43], Singh et al. [45], Mishra [15], EdNet [38], Sharma et al. [39], and Amit et al. [40]. Most of these cover more or less the same ground or discuss similar points. So we will not repeat them here, the reader may check the references. Rather, we will present a few representative impact points from NPTEL.

### 4.1 Quantitative and qualitative assessment of impact of SWAYAM/NPTEL

While earlier work highlight various positive impacts of MOOCs in India (primarily SWAYAM), data and analyses on different impact factors are not available for most of its programs (or, for IITBombayX/IIMBx/mooKIT) with the exception of NPTEL. Most programs in SWAYAM (except for NPTEL) started only in 2016. So with resource mobilization, content creation, and delivery, these have been on ground for a few years only. In contrast, NPTEL is a long running program having several data points. So we appraise the NPTEL experience.

*4.1.1 Quantitative metrics of impact of SWAYAM/NPTEL*

For quantitative metrics of impact let us consider the growth of courses, enrolments, and registrations for examination (**Figure 3**). We observe healthy steady growth in each except for a brief pandemic-induced slowdown at the end of 2020. It is also interesting to analyze the reasons behind choosing NPTEL (MOOCs). Learners of NPTEL are asked the reason/s for taking the courses and are given six options. The summary of the learners' responses over five semesters from July 2020 to July 2022 are given in **Figure 4** and are ranked from high to low:

1. To earn credit

2. To learn and update knowledge

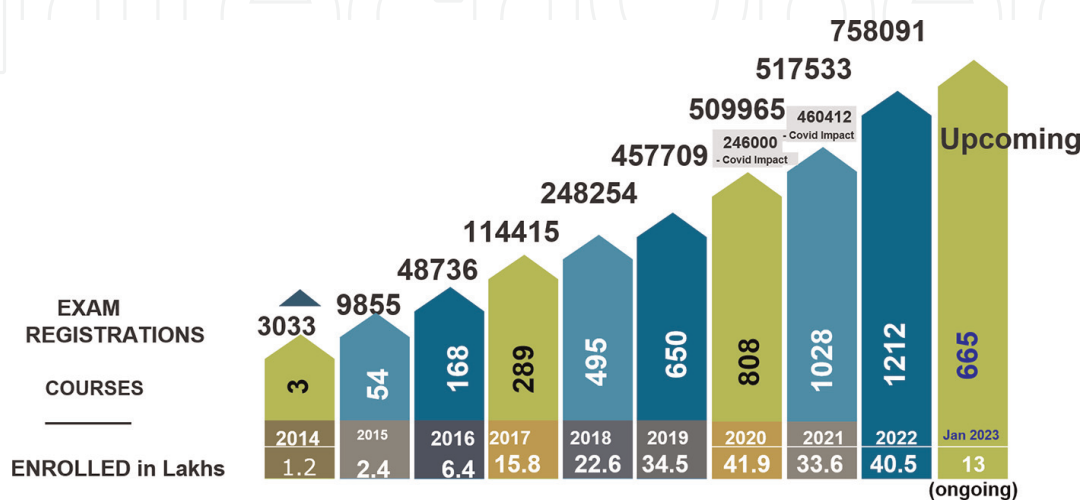

**Figure 3.**
*Growth of NPTEL course offering: 2014–2022.*

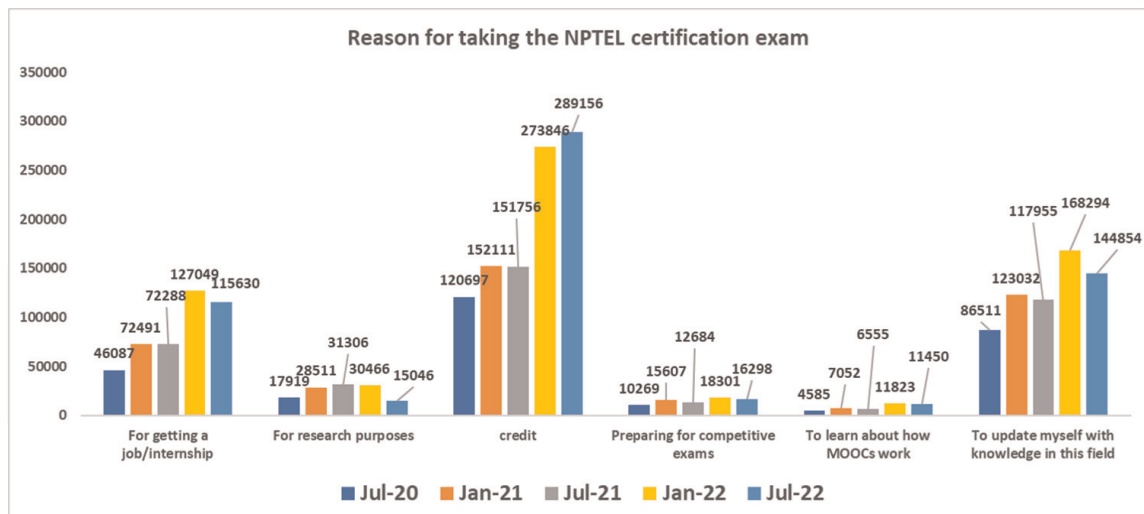

**Figure 4.**
*Why do students study NPTEL courses and take certification examination?*

3. To get job/internship

4. To empower for research

5. To prepare for competitive examination

6. To understand how MOOCs work

So learners choose MOOCs for credit mobility (compensate for non-available or low-quality courses in parent institutes), to update knowledge (possibly driven by a large group of learners who are teachers of different colleges), and for better placement opportunity (that is, industry-appropriate skilling). Contrary to expectations, preparing for competitive examination (being NPTEL, the only major examination is GATE which is also taken for job opportunity in government) is a less favored reason compared to empowerment for research.

### 4.1.2 Qualitative observations on impact of SWAYAM/NPTEL

There have been some qualitative studies on the impact of NPTEL. Notably, a white paper from the Department of Humanities and Social Sciences, IIT Madras [140] and an interview with Prof. Andrew Thangaraj [141]. In summary:

- Better education and learning

  - 80% of the enrolment through NPTEL local chapters in 4000 colleges

  - Fills in gaps for lack of good teachers

  - Students and colleges value the certification from the IITs

  - 82.9% colleges feel students gained confidence on technical interviews & technical assessments

- Employment focus of NPTEL

    ○ Recognizes active learners as NPTEL stars

    ○ Provides free soft skills training to course toppers and stars

    ○ Conducts: Employability assessment, Online mock interviews, Live video sessions, and Personalized feedback session for improvement

    ○ Exam Preparatory: "*NPTEL videos have been greatly helpful for cracking examinations such as the NET and GATE*"—Mr. Sandeep Kumar, an Assistant Professor in Electronics and Communication in PSIT Kanpur

We do observe a strong alignment between the qualitative and qualitative assessment of impact.

### 4.2 Quora speak

The assessment of impact presented above are based on data from NPTEL team or analysis carried out with the direct involvement of NPTEL team members. So we wanted to explore independent public perspective from the Internet. For this we searched for questions asked on Quora about the credibility of NPTEL and the value of certificates earned from NPTEL. We observed several positive trails on a variety of questions. As indicative samples, we quote two responses each from two questions below:

- Are NPTEL certificates good for a job and resume?[39]

- S Chetan Pandit, Tech Evangelist, IIT Graduate, Works for Qualcomm. Sep 16, 2020

    *"I am in high tech industry for more than a decade now during this time I have worked with several top MNCs, done many fresh graduates (IITs/NITs/BITS/DCEs etc) campus hiring and also experienced (lateral) hirings. so let me give you a hiring manage perspective.*
    *As an hiring manager, you would only look for skillset and gauge if the candidate has the relevant knowledge of the field, aptitude and softskills to fit in your engineering team and deliver.*
    *If all matches then he/she is good to hire. therefore, look at NPTEL courses as an opportunity to acquire new knowledge (primarily), for sure certificate is also helpful but mainly to get you shortlisted and bring you to the interview table, beyond that only learning/knowledge will help. so according to me,*
    *JOB = 20% Certificate + 80% Knowledge (learnt during Course)"*

- Abhinav Ankit, Entrepreneur, Trader, Investor, YouTuber. Oct 11, 2020

    *"Yes, of course, yaar it is NPTEL certificate. It adds an uncomparable value to your life.*

---

[39] https://www.quora.com/Are-NPTEL-certificates-good-for-a-job-and-resume

> *But yes, This is absolutely right that you will not get the job only on this certificate but it 100% adds a special value in your resume and due to that you will get job first in comparison to your competitors for a particular job.*
> *Ok, you do one thing, that I am leaving here one video link in which there is complete knowledge given about NPTEL and about its certification program and its value addition in any individuals life and this video contains every information which you want to know about NPEL.*
> *So, You go through this video [142] and get rid of your all questions and confusions."*

- Do certificates from NPTEL courses hold any value?[40]

- Ashutosh Singh, Subject Matter Expert in CSE. Nov 19, 2021.

> *"Yes, if your score is more than 90%, you become very desirable.*
> *It is relatively easy to get certified by NPTEL (scoring the minimum marks required for getting the certificate), but it is very tough to get more than 90% over all score. My highest score till date is only 83%.*
> *Marks matter a lot! When it's a matter of recruitment. A GPA of 4.5 from a tier-II engineering college is better than a GPA of 3.5 from a tier-I college.*
> *Till date (starting from 2016) I have done 7 certificate courses in the discipline of Computer Science & Engineering from NPTEL (totaling to 60 weeks of course duration), for which I have recently been recognized and honored as NPTEL Discipline Star for Dec 2019.*
> *… (images of certificates)*
> *SWAYAM-NPTEL is perhaps the best thing done in Indian education system so far. Providing free and fair access to world class education to anyone, anywhere, who wants to learn."*

- Yogesh SP, Product Evangelist at Guvi. Oct 5, 2018

> *"I am giving my perspective with respect to NPTEL certificates.*
> *Currently, I had 2 NPTEL certificates one for IMAD and another one for Data Mining.*
> *I am also appearing for two more exams. I had worked for MNC's and startups before. The interviews I had seen considers skills more than your certificates.*
> *Learning is a continuous process and it doesn't end with a certificate.*
> NPTEL is a great learning ground so make use of it in your college days."

A search for negative responses on NPTEL did not produce any credible and consistent trail.

## 4.3 Benefit to students, teachers, and colleges

The key benefits accrued to students, teachers, and colleges are:

- Students are able to learn from good teachers; can take courses that are not available in their respective colleges or courses that do not have right faculty; can

---

[40] https://www.quora.com/Do-certificates-from-NPTEL-courses-hold-any-value

benefit from comprehensive learning material—presentations, videos, transcripts, and from internship and placement opportunities through industry partners

- Several teachers have been attending NPTEL courses to learn how to teach different subjects well, how to set assignments, etc.

- Colleges suffering from shortage of teachers are listing various elective and at times the core courses for credit mobility

- Economies of scale in maintaining a good domain coverage. So NPTEL can offer courses across multiple domains even for less than 10 exam takers.

- NPTEL model shows how to manage a multi-institution project without losing focus on the Open Learning agenda.

- The affordability and accessibility of higher education is tackled by MOOCs in India. SWAYAM/NPTEL is supported by GoI while IITM BS Program [65] uses corporate philanthropy (by scholarships). Both of these programs have low entry criteria (none for NPTEL and Qualifier for BS program).

*We point to various prior studies and present quantitative data and qualitative observations to show that MOOCs in India is having a significant transformational impact. This addresses* **RQ3** *(Section 1.2.1) from the output perspectives and partly from outcome perspectives. We discuss more about outcome in the next section.*

## 5. Roadmap for MOOCs in India

While announcing the setting up of NDU (Section 2.6) during the Union Budget 2023–2024, Ms. Nirmala Sitharaman, Finance Minister, GoI has pronounced the country's vision on MOOCs far and firm. We quote [92]:

> "*A digital university will be established to provide access to the students across the country for world-class quality universal education with a personalised learning experience at their doorsteps.*"

We have long and strong tradition in MOOCs. We have been continually innovating to customize MOOCs and strengthen the ecosystem. We have made significant impacts. Next we need a roadmap to attain the vision and reap dividends from our demography. Understandably, there are challenges, challenges galore, on the way to the attainment. But, India has the will as well as the skills to conquer them. And the chief arsenal for this is technology—emerging technologies in which India is already one of the leaders.

We first review the recent use of technology in MOOCs and then map their use to address the imminent challenges of the space in India.

### 5.1 Emerging technologies for MOOCs

Technology has been contributing to various aspects of learning, specifically MOOCs [47], for a long time. Recently, Ahmad et al. [143] review the literature since 2014 to observe that the work on MOOCs is dominated by AI/ML (23%), Big Data (20%), Gamification (17%), Internet of Things (11%), Blockchain (9%), Metaverse (8%), and Digital Twin (7%). Mutawa et al. [144] also suggest strong integration of AI with MOOCs. Nunzio [145] feels "*MOOCs can prove to be a good business decision*," if online content can be developed more affordably using AI. So, we take a walk-through of a few AI technologies in education.

MOOCs is the business of knowledge. Hence, *Knowledge Graphs*[41] (KG) play a dominant role in its efficiency and effectiveness. Coupled with KGs, *AI-powered MOOCs* are facilitating aspects of delivery, management, and personalization.

### 5.1.1 Knowledge graph for MOOCs

Since 2007, knowledge graphs have been applied to education and many other disciplines [146]. In MOOCs, they help to solve varied range of problems including personalized recommendation, learning feedback, and learning path.

1. *Building KGs for MOOCs*: A KG for MOOCs is an organized collection of unified MOOC resources and course concepts for learners. It can be used to discover learning resource from several platforms and used for various research tasks. Consequently, several KGs [147–153] have been built in the last 5 years. The first large scale KG, *MOOC-KG*, has been built by Dang et al. [148] in 2019. It represents 28,591 instances with relations, includes 4 platforms[42], 604 universities, 18,671 teachers, and 9312 courses. In 2021, Dang et al. [150] extend this to represent five classes, 11 kinds of relations, and 52,779 entities with their corresponding properties, amounting to more than 3,00,000 triples. Notably, 24,188 concepts are extracted from text attributes of MOOCs and linked directly with corresponding Wikipedia entries. In 2019, Liao et al. [149], also applied knowledge graphing in MOOC and SPOC (Small Private Online Courses).

   - *Knowledge Extraction for KGs from transcripts and video captions*: While most KGs are built by mining text information from course description and metadata, a lot of knowledge stays embedded in multimodal sources like course outlines, transcripts, and lecture videos. Fareedah et al. [154], in 2018, identify a concept dependency graph for a MOOC through unsupervised analysis of lecture transcripts. Das et al. [155], 2019 propose a system that performs topic-wise semantic segmentation and annotation of MOOC lecture videos. Later in 2020, using semantic similarity and contextual relationship between different concepts, they analyze [156] how the instructor changes concepts during topic change. In 2021, Huang et al. [157] propose a method for automatically constructing a course KG by extracting course concepts from video captions in the context frame of the course outline, and then embedding the concepts back to the frame.

---

[41] Knowledge Graphs organize data from multiple sources, capture information about entities of interest in a given domain (like courses, lectures, learners, instructors, or platforms), and forge connections between them *via* linking and semantic metadata. KGs provide a framework for data integration, unification, analytics, and sharing.

[42] Coursera, EDX, XuetangX, and ICourse

2. *Course recommendations using KGs*: Course recommendations in MOOCs[43] is inherently more complex compared to the traditional course advisory due to the self-driven nature, high flexibility, easy availability, and diversity of MOOCs. Hence, KG-enhanced recommendation systems are appropriate for addressing the course recommendation problem for MOOCs. Zheng et al. [147] use ML in 2017 to construct a high educational KG by crawling courses information from several MOOC websites and performing entity and relation extraction to help learners study easily from MOOC courses. In 2020, Wang et al. [159] propose a GNN[44]-based *Attentional Heterogeneous Graph Convolutional Deep Knowledge Recommender* (ACKRec) that aligns learners' interests in different concepts with concepts embedded in different courses to recommend courses to learners. Chen et al. [151] propose an automated construction method for course KG in 2021 by annotating the pre-knowledge of each course and calculating the similarity between courses and study learning path recommendation algorithms. Jung et al. [160] propose a framework in 2022 for *Knowledge graph enhanced Personalized Course Recommendation* (KCPR) where internal information of MOOCs is integrated with an external knowledge base through user and course-related keywords. KCPR also uses a level embedding module that predicts the level of students and courses for better recommendation. Zhang et al. [152] propose *Knowledge Grouping Aggregation Network* (KGAN) in 2023 that uses the course graph, having relations between courses and facts, to estimate learners' potential interests automatically and iteratively.

3. *Learning Feedback and Learning Path Planning using KGs*: Besides being recommended on right courses, learners often need to assess how well are they actually learning the concepts in a course. KGs play a strong role in such learning feedback and path planning. Pan et al. [161], in 2017, propose to learn candidate concepts *via* an embedding-based method and rank them based on the learned representations. They evaluate the method using XuetangX and Coursera. Jiang et al. [162] present *MAssistant*, an interactive personal knowledge assistant for MOOC learners, in 2019. MAssistant helps users to trace the concepts they have learned in MOOCs by presenting important concepts as they watch the videos, to build their own concept graphs and explore them later. In 2023, Zhang et al. [153] design a KG with the visual search and display of knowledge points to provide learning feedback, to update feedback with the learning situation of learners, and to improve the learning efficiency through learning path planning.

4. *(Missing) KG for SWAYAM/NPTEL*: With the exception of [155, 156], none of the above work use the SWAYAM/NPTEL platform. This is surprising in a country like India which leads the world in various facets of AI/ML and emerging ICT. Interestingly, a number of initiatives like domain certification of NPTEL (Section 3.2.2) and extension courses (Section 3.6) like NPTEL GATE project, NPTEL FDP, and NPTEL++ have been built with manual analyses of course and content metadata. *It is important to build KGs of SWAYAM through automated methods and use for addressing several integration and quality questions of SWAYAM.*

---

[43] For example, the *Find your next course* service at Class Central [158] is based on simple metadata (subject, university, provider), keywords (from course titles, abstracts), filters (duration, language, level), or manually curated guides. This does not consider course interrelationships, preparedness / ambition of the learner, etc.

[44] Graph Neural Network

### 5.1.2 AI-powered MOOCs

With the increased focus on online education, researchers started to study how AI can power MOOCs to *scale with quality and yet remain financially sustainable*. Yu et al. [163] consider issues and solutions for AI-powered personalization in MOOC learning. Fauvel et al. [164] present a survey in 2018 for the goings-on in AI-powered MOOCs with the following taxonomic classification of the papers (**Table 7**).

They identify *Redefining openness in MOOCs* (to provide open infrastructure for content creators besides content consumers), *Complementing AI with human effort* (human-in-the-loop systems), and moving from *Engagement to Knowledge* as some of the priorities for research. Jordan et al. [165] also provide a nice survey of studies over 2012 to 2022: The Decade of the MOOC.

The face of AI capabilities in education (and in several other areas) has been redefined from November 2022 with launch of ChatGPT[45]. With *Generative Pre-trained Transformers* (GPT) for *Large Language Models* (LLM) being widely and publicly available, we can now build future generations of digital pedagogy with better elegance at ease. For example, Yilmaz et al. [166] present the framework for *SMIT - a Smart MOOC integrated with Intelligent Tutoring*. It is adaptive and dynamic, and is supported by learning analytics that aims to integrate *Learning Management Systems* (LMS) and *Intelligent Tutoring Systems* (ITS). A new era of convergence and integration has started. Agarwal et al. [167] have developed a system for *MOOCs for Lipreading* which can be a great aid for the hearing impaired. Recently, Khan Academy has launched *Khanmigo* [168–170] as a personalized tutor for every student, a planner and assistant for every teacher, and a writing coach for all. Several more will follow soon.

We have immense opportunity for AI-powered MOOCs to address most of the challenges[46] including *Personalized and Adaptive learning*, *Adaptive assessments*, *Automated evaluations*, *Natural Language Processing (query answering, translation, etc.)*, *Content Creation and Curation*, *Fraud Detection (plagiarism)*, *Predictive Analytics*, and *Intelligent Tutoring Systems*.

Finally, we analyze the challenges of MOOCs in India—as mentioned by researchers and and also as observed by the author as a MOOCs faculty for nearly a decade—and discuss the possible resolutions for the same using technology wherever possible.

| Learner Modeling | Improving Learning Experience | Learner Assessment |
|---|---|---|
| • Modeling Learner Engagement<br>• Modeling Learners' Knowledge | • Intelligent User Interactions<br>• Community Building | • Auto-grading<br>• Peer Grading<br>• Learning Skill Assessment |

**Table 7.**
*Taxonomic classification of the papers in AI-powered MOOCs, Fauvel et al. [164].*

---

[45] A large language model trained by OpenAI based on the GPT-3.5 architecture

[46] This list is accumulated from three regenerated responses from ChatGPT on "*How can AI improve MOOCs in India?*"

**5.2 Challenges and resolutions**

The challenges of MOOCs in India have been noted and analyzed by several authors including Chauhan et al. [139], Jaganathan et al. [41], Haumin et al. [33], Varyani et al. [44], Sagar [36], Pant et al. [43], Mishra [15], EdNet [38], Sharma et al. [39], Amit et al. [40], Aldahmani et al. [21], and Saxena et al. [171]. Many of them have suggested possible roadmaps too. We present a summary of these (and others from the author's experience and interactions) from the perspective of this chapter.

*5.2.1 Digital infrastructure for MOOCs*

The ecosystem of MOOCs relies heavily on the digital infrastructure for the learner, for the provider, and for the connectivity between the two.

***Internet Connectivity for Learners***:
A learner needs good internet connection to attend MOOCs courses.
*Roads ahead*:

- To alleviate this issue, lecture videos are made available in multiple resolutions to allow for easy downloads even over weak connectivity.

- Telecom in India is on a fast track with BharatNet [172] connecting villages at a fast rate. Its Phase 2 (of 3) is ongoing. Data services of Jio densely spreads across the length and breadth of India [173]. Thus, the number of Internet users in India is expected to grow [174] to 907 million (64% population) by 2023 from 398 million (29% population) in 2018.

***Digital Devices for Learners***:
A learner needs digital devices to attend MOOCs courses.
*Roads ahead*:

- While the availability of computers is perceived to be limited mostly to tier 1 and tier 2 cities, India has a deep penetration of mobile phones. During pandemic, India heavily used online education, even in schools, which deepened the penetration. Thus, it is expected [174] that the number of mobile users will grow to 966 million (68% of population) by 2023 from 763 million (56% population) in 2018.

- MOOCs providers should create contents in multiple form factors.

***Infrastructure for Content Creation***:
Institutions need good infrastructure comprising recording studios, trained man-power for recording and editing, and so on to create MOOCs content.
*Roads ahead*:

- SWAYAM has already taken several steps to widen the network of content creators through 9 different NCs in diverse areas. It is on its track to attain its plans for inclusion of institutions at multiple levels.

### 5.2.2 Digital pedagogy

MOOCs (and several other forms of digital, online, or hybrid education) have led to the *Digital Pedagogy*[47] where teachers create the courses, offer them to learners, and assess the performances of the learners in multiple levels of virtuality. Thus, teachers become their *digital twins*[48] under MOOCs, instructing the learners in *virtual tutoring* modes. Naturally, the success of this pedagogy depends heavily on the *adaptability of all the stakeholder* including teachers, learners, and teaching assistants; on the *quality, diversity, and multilinguality of contents* (lecture videos, transcripts, reading notes, assignments & solutions, and so on); on the *efficiency and efficacy of virtual tutoring and instructional processes*; and on the *quality of assessment, and evaluation*. MOOCs all across the world, including in India, fall shorts on all of these aspects.

**Adaptability by the Stakeholders**:

*Teachers*: The *teachers* of MOOCs record courses in the absence of the learners (lecturing merely to *virtual learners*). So most teachers, skilled in face-to-face and interactive pedagogy of delivering lectures fail to incite the excitement in their videos and often miss out to properly embed the necessary learning-by-example stubs. Further, the teachers are not trained to *look into the camera* and they get *little eye-contact* with the observer (learner). Also, most courses are recorded with presentation slides with the teacher sitting in a chair which *eliminates the communicative body language* of the teacher and makes it difficult to develop concepts step-by-step.

*Roads ahead*:

- All course creators need to abide by the *Guidelines for SWAYAM* [104], 2017. This standardizes the course plans. KGs of courses in SWAYAM and an AI-driven intelligent checker can improve compliance to the guidelines.

- While the guidelines are strong in terms of lecture and course planning, it does not talk about the presentation and communications. *Listener bots* or *bot-learners* may be created for feedback on these aspects.

- A MOOCs course on *MOOCs Pedagogy* should be created by the experts to train all teachers (could be an FDP) who offer MOOCs.

*Learners*: The *learners* of MOOCs watch the lecture videos in the absence of the teacher and the peers. They interact with the teacher, the teaching assistants, and the peers only in written form (using *course forum—a virtualized textual model of teachers, assistants, and peers*). This leads to the feeling of *isolation*, inadequate opportunity for honing skills of *verbal and interpersonal communication*, depends heavily on the *self-motivation* of the learners (which many may lack), and leads to overall *digital fatigue*[49]. *Adoption of technology* itself can be a challenge for some of the learners.

---

[47] Digital pedagogy is the study and use of contemporary digital technologies in teaching and learning. It may be applied to online, hybrid, and face-to-face learning environments.

[48] A digital twin is a digital representation of a physical object, process, service, or environment that behaves and looks like its counterpart in the real world.

[49] Digital fatigue (aka, screen burnout) is a state of mental exhaustion brought on by the excessive use of digital devices.

Mithun Mohan et al. [175] examine the key factors that influence the behavioral intention to use MOOCs among 412 university PG students. They report that 41% students feel that lack of time from academic schedule is the key barrier[50] to MOOCs, while 33% felt that MOOCs are less effective compared to classroom teaching. Technology barrier (16%) and monotonous nature of MOOCs (10%) are the other factors.

*Roads ahead*:

- SWAYAM/NPTEL address these through online course forum (for discussions with the teachers/TAs and/or peers), local chapters (SPoC in colleges to mentor students), weekly live tutoring sessions by PMRF scholars, and monthly live sessions with the teachers.

    ○ *Alarmingly these measures are becoming less effective with time*.

    ○ In Jul 2022, 96% courses in NPTEL had <100 learners on forum [103].

    ○ Most courses are getting few students on live sessions.

    ○ Pass percentage in several courses is low even for LC students.

    ○ Is digital fatigue starting to cripple MOOCs in India? In-depth research is needed to understand the learners and their learning behavior better.

    ○ Personalized bots like *Khanmigo* [169] can greatly improve engagement.

**Quality, Diversity, and Multilinguality of Contents**:

*Quality of Content*: Quality contents need highly skilled teachers, good infrastructure, and trained support man-power. At India's scale MOOCs, these are stiff barriers and solutions should constantly be sought with AI.

*Roads ahead*:

- *Guidelines for SWAYAM* [104] should be followed by all contents. Absent compliance-check or audit process[51] can be put in place using KG and AI.

- *Quality Reference Framework for MOOCs* (QRF) [58] has been released in 2018 to analyze the needs and demands of MOOCs, to design, develop, and implement new MOOCs, and to evaluate and improve existing MOOCs. It consists of three dimensions: *Phases*, *Perspectives*, and *Roles*; and provides the *QRF Key Quality Criteria* and the *QRF Quality Checklist* for designing and developing MOOCs. **SWAYAM guidelines and processes should be strengthened with the adoption of QRF**.

---

[50] It may be noted that this study was carried out in 2019 before the credit mobility was implemented across HEIs to alleviate exactly this specific issue.

[51] Digital quality guarantee processes, however, are conducted on all videos and transcriptions.

*Diversity of Content*: MOOCs in India have piggybacked on NPTEL which focused on engineering domain, relied on the expertise from IITs and IISc, and performed several techno-social innovations to bring MOOCs in India to its present state. As SWAYAM diversifies MOOCs to other domains, to broader sections of learners, and wider academic levels, the need for diverse contents (subject matter as well as form) becomes a priority. This is a huge challenge given the weak institutional bandwidth in various domain areas.

*Roads ahead*:

• A possible resolution may focus on and grow with public-private partnership in edTech some of which are already underway [89–91, 111].

• Generative AI may be used to create selective contents for some subjects.

*Multilinguality of Contents*: With 22 scheduled languages (and several more practiced ones) and deep diversity of culture, Indian MOOCs need to strongly support Indian languages besides English.

*Roads ahead*:

• As discussed in Section 3.7, NPTEL has started this translation process with the help of LCs. Nearly 300 courses in 11 languages having 40,000+ lectures for 10,000+ hours [103] are available.

• With the advances in Natural Language Processing (NLP) for Indian languages (especially Hindi), automated translation (like Google Translate), automated transcription and captioning, and Knowledge Graphs (see Section 5.1.1), accelerated automated processes should be setup for translations of the languages of the videos and transcripts.

• Beyond the (translated) language of the contents, the overall process of engagement in MOOCs needs to be made multilingual.

**Efficiency and Efficacy of Processes**:

MOOCs are created by digitally aided record-deliver-evaluate processes with focus on creating digital equivalents of physical systems of learning. If often does not ensure if the virtual tutoring and instructional processes are efficient and effective from the socio-psychological aspects of learning and learners. For example, the completion rate is taken as the metric of success and not the outcome of learning.

*Roads ahead*:

• While MOOCs advocates increased *Self Regulated Learning* (SRL) capability of a learner, the NPTEL model shows how processes (Weekly Release, Deadlines, Grading that permits flexibility) along with group/external regulation helps in increasing uptake of online courses.

• "*By strategically aligning the technology and processes, we can really attempt to solve the larger issues of learner diversity and sustainability (of learning) at scale.*" – Dr. Jayakrishnan M, Senior Scientist, NPTEL

- *Stickiness of learners*[52] (at ~85% retention) is ensured in IITM BS program [65] with a team of instructors and a dedicated operations team.

- Deep research in digital pedagogy and use of AI are needed to obtain the desired outcomes (Section 5.1.2).

- MoE has launched NEAT [90] as a PPP model (Section 3.8.1) to showcase Technology Products using AI for customized learning or e-content in niche areas having highly employable skills.

### *Quality of Assessment and Evaluation*:

Assessment and evaluation is one of the weak areas of MOOCs (and many forms of online education). The assignments and examinations are based on Multiple Choice (MCQ), Multiple Select (MSQ), Short Answer (SA), and Fill-in-the-blank questions. This limits the ability to test the breadth and depth of the understanding of the learners. For example, in a programming language course offered by the author for over last 8 years, actual programming assignments could never be part of the evaluation process. This also is a major hindrance for various social science courses including law where descriptive answers are a necessity as a learning measure. Overall, MOOCs provide only a few assessments and little feedback. Hence, it is not possible to assess knowledge for cognitive skills beyond mere information retrieval or to handle abstraction or to take complex decisions, and so on. Marks scored in assignments have very low credibility due to rampant plagiarism[53]. Hence, proctored physical examinations are needed for credible evaluation. This limits the scaling of MOOCs and with only one such proctored examination in a course, the students do not get the required learning practice.

*Roads ahead*:

- Major MOOCs platform use multiple technology-driven strategies to address the problem of plagiarism. These include remote proctoring by humans invigilators (or smart machines?), and use access to microphone and camera of student's computer to make random recordings during examinations (and screen using surveillance software) during exams. IoT may provide some solutions in future.

- Research and interventions are needed to improve the processes of semi-automated/automated evaluation to scale MOOCs with quality.

- *Robust, complete, scalable, and flexible* processes for assessment and evaluation are possible today with AI (Section 5.1.2). LLMs with course KGs and pre-trained transformers may be built to evaluate subjective answers and provide feedback.

**Digital Pedagogy in India is still in its infancy. And success of MOOCs in the longer run and national scale critically depends on it**.

---

[52] Learners taking courses consistently in subsequent semesters after the initial exposure

[53] As the NPTEL team observed in [103], in July 2022 courses, many students had high average assignment marks scored only less than 10% marks in final examination.

### 5.2.3 MOOCs in principle—Not on paper

**Retention/Completion Rate**:

MOOCs in India usually have high enrolment but low retention/completion of the courses. In NPTEL only about 10% of the students (**Figure 3**) who enroll actually go on to complete the course [103, 120]. In 500+ SWAYAM courses through the CEC, UGC, and IGNOU till 2019, only 30–40% students who took the examination actually passed and completed the course (**Table 8** in [43]). While this is a global phenomenon (e.g., Chiappe and Castillo [176] report completion rate of MOOCs between 5 and 15%), India needs to address it effectively to improve the effectiveness of its investments. *At 10% does MOOCs in India remain Massive or is it simply Large?*

At this point, may we question how should the retention/completion rate be estimated? Five distinct measure points have usually been considered by the researchers. We compute various ratio metrics using the data of NPTEL [103, 120] during Jan-Apr 2019 to Jan-Apr 2022 in **Table 8**.

There are three large losses in the table: (a) *nEnrol → nOneWeek* = 65%, (b) *nOneWeek → nRegister* = 65%, and (c) *nAppear → nFail* = 25%. (a) may be ignored—it is the check-out time expected in any open flexible system. These are students who did not even get engaged to complete the first assignment. (c) is not alarmingly high. It can improve only with the quality of the courses and offerings—a slow process. (b) needs specific attention:

| Metric | % | | Remarks |
|---|---|---|---|
| nOneWeek/ nEnrol | = | 35% | • 65% students leave within a week |
| nRegister/ nOneWeek | = | 35% | • Further 65% students leave between 2nd week & registration |
| nRegister/ nEnrol | = | 12% | • 88% students leave between enrolment & registration |
| nAppear/ nRegister | = | 85% | • 15% students register, but do not appear. This is higher, yet commensurate with non-MOOC courses |
| nPass/nAppear | = | 75% | • 25% failure rate is not particularly high |
| nPass/nEnrol | = | 8% | • *This is alarming* |
| nPass/ nOneWeek | = | 22% | • This is better, yet too low |
| Typical metrics or measure points as used in Table 8 in [43, 103, 120, 176] | | | |
| nEnrol | = | | # of students enrolled |
| nOneWeek | = | | # of students who completed the first assignment, that is, 1 week |
| nRegister | = | | # of students who registered for the final examination |
| nAppear | = | | # of students who appeared for the final examination |
| nPass | = | | # of students who passed the course |
| nFail | = | | # of students who failed the course = *nAppear – nPass* |

**Table 8.**
*Retention rate for NPTEL: 2019–2022.*

- These students left the course at some point after completing the first assignment. So did they take the decision to discontinue after getting the evaluation of the first assignment? To understand this, it is necessary to track how many of them submitted the second assignment. Those students should also be clubbed with (a) and ignored.

- So the students who left the course after the second (or third, or later) assignment/s, may have done so for the reasons outlined in Section 5.2.2 in terms of various shortcomings of the digital pedagogy including lack of engagement with the teacher, lack of self-motivation, lack of English language, skills, digital fatigue, and so on. We need to know.

*Roads ahead*:

- An in-depth analysis for the case (b) for SWAYAM/NPTEL is necessary.

- Declining enrollments in MOOCs worldwide is leading to MOOCs 2.0 [43] to explore freemium[54] models (like NPTEL+), offering a mix of free and paid courses. MOOCs in India should check out as well.

- Several innovations mentioned in Section 3 including interactive forums, local centers, and live sessions are targeting to improve retention.

- In the IITM BS program [65], a team of instructors along with a very capable operations team helps to ensure near 85% retention. It is exploring Peer-Learning and Peer-Support models for sustenance.

- AI-driven solutions discussed in Section 5.1.2 including personalized tutor, TA-bots, and explainer tutorials can improve retention. For example, Sharma [117] apply *learning analytics* on NPTEL data to build a model to predict dropouts (and hence, design measures to reduce it). Dalipi et al. [60] use *sentiment analysis* of students' feedback on MOOCs for understanding dropouts. More such studies are needed.

### Learning Outcomes:

  While under the *Guidelines for SWAYAM* [104], every course needs to specify the learning outcomes of the course *a priori*, there has not been any study on the a posteriori learning outcomes for the finishers of the courses. In a related question, Sharma et al. [39] asks for a direct comparison of MOOCs credits to classroom credits. She argues if two students earn the same set of credits—one from an institution with payment and the other from NPTEL-NOC—will the former be able to differentiate herself from the latter. If not, *Tragedy of the Commons*[55] will occur where, moving forward, no learner will be willing to pay for the course anymore. Singh and

---

[54] Freemium is a business model in which a company offers basic or limited features to users at no cost and then charges a premium for supplemental or advanced features.

[55] The tragedy of the commons refers to a situation in which individuals with access to a public resource (called a common) act in their own interest and, in doing so, ultimately deplete the resource.

Kakkar [124, 125] show that while credit mobility draws more students to SWAYAM, the certification rate is poor and SWAYAM fails in its desired outcome goals.

*Roads ahead*:

- Specific studies should be instituted to understand the learning outcomes of various SWAYAM/NPTEL courses in different domains.

- KGs of SWAYAM courses should be build with models of learning outcomes by concepts for a detailed analysis to improve outcomes. Employment history should also be tracked in this regard.

*Gender Gap*:

Sharma et al. [39] report gender disparities in the enrollment rate of MOOCs by quoting other researchers. However, no data is provided to support the claim. An indirect indicator of this gap can be observed in NPTEL [103] where since 2016, 15–25% more male students have registered for examinations compared to female students. *So, while MOOCs in India is Open by structure, is it only partially so under societal dynamics?*

*Roads ahead*:

- Specific studies should be instituted to understand the gender disparity, if any, in various SWAYAM/NPTEL courses in different domains.

Thus, restructuring of the MOOCs platforms, using emerging technology for better ecosystem, is needed to match the changing needs and requirements of our students' present and future demands.

*We have presented a brief survey of key emerging technologies that can immediately benefit MOOCs. We then explore various challenges of MOOCs in India with quantitative data and qualitative observations. For every challenge we outline possible resolutions using adoption of technology, processes, standardization, collaborations, and innovations. This addresses* **RQ4** *(Section 1.2.1). Further, we stitch the vision with the challenges and resolutions in a flow to present a roadmap for MOOCs in India for a decade and address* **RQ5** *(Section 1.2.1).*

## 6. Conclusions

We have taken a deep look into MOOCs in India starting with an evolutionary review of the last 50 years (1974–2023) on how MOOCs not only started and grew, but is fast becoming a leading education system in India. This evolution and emergence of strength has been possible due to a number of unique innovations that have resulted in huge impacts—mostly positive—which we have assessed from primary and secondary sources. We have also elucidated the challenges with pointers to possible resolutions from various studies by researchers coupled with the first-hand experience of the author with multiple MOOCs. We finally outline a roadmap with the immense possibilities of development and growth in both quantitative and qualitative terms using AI-powered MOOCs.

Pandemic catalyzed rapid growth and adoption in the space of MOOCs. While schools and colleges shut down for several months, India continued to learn in domains where MOOCs had presence. Sagar [36] argued in 2021 that MOOCs would be the future of education in India. As open and free models of MOOCs started to crumble in parts of the world with major players like Coursera and edX turning commercial [178], SWAYAM stood a ground of strength with the escalated support of NEP 2020 at a national scale. In 2022, Shah et al. [118] predicted a powerful wave as 60+ Indian Universities started offering about 500 degrees online degrees and Amit et al. [40] emphasized on the need of deeper use of technology for engaging academic materials through animation, illustrations, and video descriptions to MOOCs reach to its potential in target markets.

The most recent shot in the arm has been wider public availability of powerful LLMs and GPTs over the last few months with private free players like Khan Academy [168–170] coming out with innovative personalization of online learning. India has inherent strengths in software, AI, innovation, and start-ups. With focused research

What is the future of MOOCs in India for the next decade?

MOOCs (Massive Open Online Courses) have gained popularity in India in recent years, and their future prospects look promising. India has one of the largest populations of students globally, and the government has also been focusing on promoting online learning in the country. Here are some of the potential trends for MOOCs in India over the next decade:

1. Increase in the number of users: With the increasing internet penetration in the country, the number of users accessing MOOCs is expected to increase significantly. The government's push towards digitization and the availability of affordable smartphones and internet connectivity will likely make MOOCs accessible to a larger audience.
2. Expansion of courses: MOOCs in India are expected to expand beyond traditional subjects and include courses on emerging technologies, management, and vocational skills. This diversification of courses will cater to a wider audience and help bridge the skill gap in the country.
3. Blended learning: While MOOCs are popular for their flexibility, they are also criticized for their lack of personalized learning experience. To overcome this, MOOCs are expected to adopt blended learning, where students can combine online learning with traditional classroom learning.
4. Corporate training: MOOCs are also expected to expand their reach to corporate training. With businesses looking for cost-effective and flexible training solutions, MOOCs can provide training on various skills, including leadership, management, and technology.
5. Collaboration: MOOCs are expected to collaborate with universities and other institutions to offer credit-bearing courses. This collaboration will help students get recognized degrees and certificates from reputed institutions.

In summary, the future of MOOCs in India looks bright, and they are expected to play a significant role in transforming the education and training landscape in the country.

**Figure 5.**
*Responses from ChatGPT on "what is the future of MOOCs in India for the next decade?"*

to power MOOCs with AI, India can alleviate several of the current challenges within the next 5 to 10 years. Thus, it can emerge as a leader in the balanced blend of MOOCs and traditional systems of education for its young global workforce.

Before we close, let us note what ChatGPT has to say on "*What is the future of MOOCs in India for the next decade?*" (**Figure 5**). Ironically, it humbly does not (cannot) predict the strong role it itself is expected to play in the process.

## Acknowledgements

The author deeply appreciates the support of Dr. Jayakrishnan M, Senior Scientist, NPTEL, Ms. Bharathi Balaji, Senior Project Officer, NPTEL, Mr. Shib Shankar Das, Senior Manager, NPTEL and Prof. Andrew Thangaraj, Professor, IIT Madras and Principal Investigator, NPTEL for sharing data on NPTEL and for providing continuous support in the development of this chapter through in-depth discussions. He is also indebted to Dr. Jayakrishnan and Mr. Srijan Bandyopadhyay, Course Development Manager, AshokaX, Ashoka University for painstakingly reviewing earlier versions of this chapter.

## Abbreviations

| | |
|---|---|
| ABC | Academic Bank of Credits |
| AI | Artificial Intelligence |
| AICTE | All India Council for Technical Education |
| CBSE | Central Board of Secondary Education |
| CBT | Computer-Based Test |
| CC | Class Central |
| CEC | Consortium For Educational Communication |
| CET | Centre for Education Technology |
| CII | Confederation of Indian Industry |
| CMU | Carnegie Mellon University |
| DIKSHA | Digital Infrastructure for Knowledge Sharing |
| DL | Deep Learning (in AI) |
| DL | Digital Library (in Library and Information Sciences) |
| DTH | Direct To Home |
| ERNET | Education and Research Network |
| ELNet-3 L | Electronically Networked – Life Long Learning |
| ET | The Economic Times |
| FICCI | Federation of Indian Chambers of Commerce & Industry |
| FOSS | Free and Open-Source Software |
| FOSEE | Free/Libre and Open Source Software for Education |
| GATE | Graduate Aptitude Test in Engineering |
| GER | Gross Enrolment Ratio |
| GIAN | Global Initiative of Academic Networks |
| GoI | Government of India |
| GPT | Generative Pre-trained Transformers |
| HE/HEI | Higher Education/Higher Education Institution |
| ICT | Information and Communication Technology |
| IGNOU | Indira Gandhi National Open University |

| IIT | Indian Institute of Technology – IITM (Madras), IITKGP (Kharagpur), IITB (Bombay), IITD (Delhi), IITK (Kanpur), IITG (Guwahati), IITR (Roorkee) |
| --- | --- |
| IIM | Indian Institute of Management – IIMA (Ahmedabad), IIMB (Bangalore), IIMC (Calcutta), IIML (Lucknow) |
| INFLIBNET | Information and Library Network Centre |
| ITS | Intelligent Tutoring Systems |
| JNU | Jawaharlal Nehru University |
| KG | Knowledge Graph |
| LLM | Large Language Models |
| LMS | Learning Management System |
| MCQ | Multiple Choice Questions |
| MSQ | Multiple Select Questions |
| MHRD | Ministry of Human Resource Development |
| ML | Machine Learning |
| MoE | Ministry of Education |
| MOOC | Massive Open Online Course |
| NAD | National Academic Depository |
| NCERT | National Council of Educational Research and Training |
| NDLI | National Digital Library of India |
| NDU | National Digital University |
| NEAT | National Education Alliance for Technology |
| NETF | National Educational Technology Forum |
| NEP | National Education Policy |
| NIOS | (The) National Institute of Open Schooling |
| NITTTR | National Institute of Technical Teachers' Training and Research |
| NKN | National Knowledge Network |
| NLP | Natural Language Processing |
| NMEICT | National Mission on Education through Information and Communication Technology |
| NOC | NPTEL Online Certification |
| NPTEL | National program on Technology Enhanced Learning |
| OER | Open Educational Resource |
| PDS | Plagiarism Detection Software |
| PIB | Press Information Bureau |
| PTI | Press Trust of India |
| PPP | Public-Private Partnerships |
| SPoC | Single Point of Contact (in NPTEL) |
| SPOC | Small Private Online Courses |
| SLM | Self Learning Material |
| STEM | Science, Technology, Engineering, and Mathematics |
| SWAYAM | Study Webs of Active–learning for Young Aspiring Minds |
| TLP | Teaching-Learning Process |
| ToI | Times of India |
| UGC | University Grants Commission |
| V-C | Vice-Chancellor |
| VC | Video Conference |
| VCTEL | Virtual Centre for Technology Enhanced Learning |
| VSNL | Videsh Sanchar Nigam Limited |
| WOTEL | Workshop On Technology Enhanced Learning |

## Author details

Partha Pratim Das[1,2]

1 Department of Computer Science and Center for Data Science and Analytics, Ashoka University, Sonepat, India

2 Department of Computer Science and Engineering, Indian Institute of Technology Kharagpur, Kharagpur, India

*Address all correspondence to: ppd@ashoka.edu.in; ppd@cse.iitkgp.ac.in; partha.p.das@gmail.com

IntechOpen